\documentclass[acmsmall]{acmart}
\AtBeginDocument{%
  }


\setcopyright{rightsretained}
\acmDOI{10.1145/3643734}
\acmYear{2024}
\copyrightyear{2024}
\acmSubmissionID{fse24main-p79-p}
\acmJournal{PACMSE}
\acmVolume{1}
\acmNumber{FSE}
\acmArticle{8}
\acmMonth{7}
\received{2023-09-27}
\received[accepted]{2024-01-23}

\usepackage{amsmath}  

\usepackage[ruled,linesnumbered]{algorithm2e}

\usepackage{multirow}
\usepackage{algpseudocode}
\usepackage{tikz}
\newcommand{\circnumber}[1]{\lower.75ex\hbox{\tikz\draw (0pt, 0pt)%
    circle (.47em) node {\makebox[.15em][c]{\small #1}};}}
\newcommand{\ballnumber}[1]{\lower.75ex\hbox{\tikz\fill(0pt, 0pt)%
    circle (.5em) node {\makebox[.15em][c]{\small \textcolor{white}{#1}}};}}
\newcommand{\dcircnumber}[1]{\lower.75ex\hbox{\tikz\draw(0pt, 0pt)%
    circle (.47em) circle (.37em) node {\makebox[.15em][c]{\small #1}};}}

\newcommand{\emptycirc}[1]{\lower.75ex\hbox{\tikz\draw (0pt, 0pt)%
    circle (.47em) node {\makebox[.15em][c]{\small \textcolor{white}{#1}}};}}
\newcommand{\emptyball}[1]{\lower.75ex\hbox{\tikz\fill(0pt, 0pt)%
    circle (.5em) node {\makebox[.15em][c]{\small #1}};}}
\newcommand{\emptydcirc}[1]{\lower.75ex\hbox{\tikz\draw(0pt, 0pt)%
    circle (.47em) circle (.37em) node {\makebox[.15em][c]{\small \textcolor{white}{#1}}};}}

\usepackage{threeparttable} 
\usepackage{listings, xcolor}
\usepackage{enumitem}
\definecolor{verylightgray}{rgb}{.97,.97,.97}
\lstdefinelanguage{Solidity}{
  keywords=[1]{anonymous, assembly, assert, balance, break, call, callcode, case, catch, class, constant, continue, constructor, contract, debugger, default, delegatecall, delete, do, else, emit, event, experimental, export, external, false, finally, for, function, gas, if, implements, import, in, indexed, instanceof, interface, internal, is, length, library, log0, log1, log2, log3, log4, memory, modifier, new, payable, pragma, private, protected, public, pure, push, require, return, returns, revert, selfdestruct, send, solidity, storage, struct, suicide, super, switch, then, this, throw, transfer, true, try, typeof, using, value, view, while, with, addmod, ecrecover, keccak256, mulmod, ripemd160, sha256, sha3}, 
  keywordstyle=[1]\color{blue}\bfseries,
  keywords=[2]{address, bool, byte, bytes, bytes1, bytes2, bytes3, bytes4, bytes5, bytes6, bytes7, bytes8, bytes9, bytes10, bytes11, bytes12, bytes13, bytes14, bytes15, bytes16, bytes17, bytes18, bytes19, bytes20, bytes21, bytes22, bytes23, bytes24, bytes25, bytes26, bytes27, bytes28, bytes29, bytes30, bytes31, bytes32, enum, int, int8, int16, int24, int32, int40, int48, int56, int64, int72, int80, int88, int96, int104, int112, int120, int128, int136, int144, int152, int160, int168, int176, int184, int192, int200, int208, int216, int224, int232, int240, int248, int256, mapping, string, uint, uint8, uint16, uint24, uint32, uint40, uint48, uint56, uint64, uint72, uint80, uint88, uint96, uint104, uint112, uint120, uint128, uint136, uint144, uint152, uint160, uint168, uint176, uint184, uint192, uint200, uint208, uint216, uint224, uint232, uint240, uint248, uint256, var, void, ether, finney, szabo, wei, days, hours, minutes, seconds, weeks, years},  
  keywordstyle=[2]\color{teal}\bfseries,
  keywords=[3]{block, blockhash, coinbase, difficulty, gaslimit, number, timestamp, msg, data, gas, sender, sig, value, now, tx, gasprice, origin},  
  keywordstyle=[3]\color{violet}\bfseries,
  identifierstyle=\color{black},
  sensitive=false,
  comment=[l]{//},
  morecomment=[s]{/*}{*/},
  commentstyle=\color{gray}\ttfamily,
  stringstyle=\color{red}\ttfamily,
  morestring=[b]',
  morestring=[b]"
}
\begin{document}



\title{Efficiently Detecting Reentrancy Vulnerabilities in Complex Smart Contracts}

\author{Zexu Wang}
\orcid{0009-0004-1439-2989}
\affiliation{%
  \institution{Sun Yat-sen University}
  \city{Zhuhai}
  \country{China}
}
\affiliation{%
  \institution{Peng Cheng Laboratory}
  \city{Shenzhen}
  \country{China}
}
\email{wangzx97@mail2.sysu.edu.cn}

\author{Jiachi Chen}
\orcid{0000-0002-0192-9992}
\affiliation{%
  \institution{Sun Yat-sen University}
  \city{Zhuhai}
  \country{China}
}
\email{chenjch86@mail.sysu.edu.cn}

\author{Yanlin Wang}
\orcid{0000-0001-7761-7269}
\affiliation{%
  \institution{Sun Yat-sen University}
  \city{Zhuhai}
  \country{China}
}
\email{wangylin36@mail.sysu.edu.cn}

\author{Yu Zhang}
\orcid{0000-0003-2040-5059}
\affiliation{%
  \institution{Harbin Institute of Technology}
  \city{Harbin}
  \country{China}
}
\affiliation{%
  \institution{Peng Cheng Laboratory}
  \city{Shenzhen}
  \country{China}
}
\email{yuzhang@hit.edu.cn}

\author{Weizhe Zhang}
\orcid{0000-0003-4783-876X}
\affiliation{%
  \institution{Harbin Institute of Technology}
  \city{Harbin}
  \country{China}
}
\affiliation{%
  \institution{Peng Cheng Laboratory}
  \city{Shenzhen}
  \country{China}
}
\email{wzzhang@hit.edu.cn}

\author{Zibin Zheng}
\authornote{Corresponding Author}
\orcid{0000-0002-7878-4330}
\affiliation{%
  \institution{Sun Yat-sen University}
  \city{Zhuhai}
  \country{China}
}
\affiliation{%
  \institution{GuangDong Engineering Technology Research Center of Blockchain}
  \city{Zhuhai}
  \country{China}
}
\email{zhzibin@mail.sysu.edu.cn}


\begin{abstract}
Reentrancy vulnerability as one of the most notorious vulnerabilities, has been a prominent topic in smart contract security research. Research shows that existing vulnerability detection presents a range of challenges, especially as smart contracts continue to increase in complexity. Existing tools perform poorly in terms of efficiency and successful detection rates for vulnerabilities in complex contracts.

To effectively detect reentrancy vulnerabilities in contracts with complex logic, we propose a tool named SliSE. SliSE's detection process consists of two stages: \textit{Warning Search} and \textit{Symbolic Execution Verification}. In Stage \uppercase\expandafter{\romannumeral1}, SliSE utilizes program slicing to analyze the \textit{Inter-contract Program Dependency Graph} (I-PDG) of the contract, and collects suspicious vulnerability information as warnings. In Stage \uppercase\expandafter{\romannumeral2}, symbolic execution is employed to verify the reachability of these warnings, thereby enhancing vulnerability detection accuracy.
 SliSE obtained the best performance compared with eight state-of-the-art detection tools. It achieved an F1 score of 78.65\%, surpassing the highest score recorded by an existing tool of 9.26\%. Additionally, it attained a recall rate exceeding 90\% for detection of contracts on Ethereum. Overall, SliSE provides a robust and efficient method for detection of Reentrancy vulnerabilities for complex contracts.
    
\end{abstract}




\maketitle

\section{Introduction}
\label{sec:intro}

As decentralized applications (DApps) become more versatile in functionality, the complexity of the underlying contract logic has correspondingly increased. This escalation in complexity poses significant challenges for existing tools in detecting vulnerabilities within complex contracts. Reentrancy vulnerabilities are one of the most notorious types~\cite{swe}. Starting from the DAO Reentrancy attack~\cite{DAO} in 2016 that caused a \$150 million loss in digital assets, Reentrancy attacks on blockchain continue to occur. Concurrently, many academic studies and tools for detecting Reentrancy vulnerabilities have emerged. 


Various approaches have been employed to detect Reentrancy vulnerabilities, including symbolic execution~\cite{29mossberg2019manticore,luu2016making,24zhang2019mpro,13torres2018osiris,9mythril}, fuzz testing~\cite{choi2021smartian,jiang2018contractfuzzer,nguyen2020sfuzz,so2021smartest}, static analysis~\cite{11ye2020clairvoyance,feist2019slither,bose2022sailfish}, and formal verification ~\cite{contro2021ethersolve,tsankov2018securify,feist2019slither}. However, most existing tools have been evaluated on relatively simple contract datasets, e.g., the \textit{SmartBugs Dataset}~\cite{8Smartbugs}, lacking experimental assessments on real-world complex contracts. To ascertain their efficacy in complex DApps, Zheng et al.~\cite{zheng2023dappscan} curated a dataset that comprises 895 vulnerabilities. These vulnerabilities were obtained from 1,322 open-source DApp audit reports provided by 30 blockchain security companies, covering 25 types of vulnerabilities. Compared to the \textit{SmartBugs dataset}~\cite{8Smartbugs}, it has approximately 25 times the average lines of code and 30 times the average function count. The study also evaluated five state-of-the-art vulnerability detection tools~\cite{bose2022sailfish,choi2021smartian,9mythril,tsankov2018securify,feist2019slither} on this dataset. The experimental results revealed that most tools had a low success detection rate (less than 30\%), especially for Reentrancy vulnerabilities (less than 11\%). This emphasizes the necessity of focusing on detecting real-world vulnerabilities in complex contracts rather than simple toy contracts in future research.


Real-world Reentrancy attack events usually involve complex function call relationships. However, most existing tools focus only on the security of a single function within the contract, which is insufficient to guarantee the overall contract safety. Compositional Reentrancy vulnerabilities of smart contracts are introduced by cross-contract interactions, such as cross-function Reentrancy~\cite{cross_function_Re} and cross-contract Reentrancy~\cite{cross_contract_Re}, presenting considerable challenges. Effective detection of compositional Reentrancy vulnerabilities requires a thorough analysis of contract interactions and a systematic examination of data and control flow transitions. Incorporating program semantics could enhance the accuracy of identifying critical data and control flows, thereby improving compositional Reentrancy vulnerability detection effectiveness. Additionally, the computation of dynamic jump addresses poses another major challenge, often resulting in incomplete Control Flow Graph (CFG) paths. Despite complex function call relationships may generate numerous execution paths, many current tools predominantly rely on static stack simulation for path recovery. This method struggles with dynamic jump addresses, leaving CFG paths incomplete and compromising the accuracy of symbolic execution. Therefore,  the analysis of compositional Reentrancy vulnerability and CFG path recovery are the key challenges in complex contract Reentrancy vulnerability detection.

To address these challenges, we propose the SliSE method, which combines symbolic execution with program slicing to detect Reentrancy vulnerabilities in complex contracts. The detection process is divided into two stages: \textit{Warning Search} and \textit{Symbolic Execution Verification}. In the \textit{Warning Search} stage, program slicing is performed based on program dependencies to search and extract critical paths. Subsequently, in the \textit{Symbolic Execution Verification} stage, the reachability of these critical paths is further verified to achieve efficient Reentrancy vulnerability detection. Through program dependency analysis, the I-PDG (Inter-contract Program Dependency Graph) of contract is constructed, which can provide cross-contract data and control dependencies. Combined with our Reentrancy vulnerability slicing standards, the I-PDG is sliced to prune paths. SliSE analyzes program dependency among critical instruction statements (such as \textit{require}, \textit{assert}, etc.) to model the semantics, then identifies data and control flow transitions. With this information, it verifies whether these semantics adhere to the secure C-E-I (\textit{Check->Effect->Interaction}) patten. This pattern mandates timely state updates before interacting with external contracts, maintaining atomicity of transactions and prevent Reentrancy (see section~\ref{sec:bkg&mot} for details). If deviations from this pattern are detected, SliSE generates a warning that includes the corresponding function and location information. In the \textit{Symbolic Execution Verification} stage, SliSE employs the CFG path recovery algorithm to recovery execution paths and gather essential path constraints. Finally, it proves the reachability of the path and the existence of vulnerabilities through constraint solving.


We evaluated SliSE's performance by comparing the detection results with eight state-of-the-art tools, analyzing its performance in detecting Reentrancy vulnerabilities from complex contracts. The results show that SliSE performs well in detecting complex contract Reentrancy vulnerabilities, with the F1 score of 78.65\%, significantly exceeding the highest score of 9.26\% achieved by the eight state-of-the-art tools. To assess its effectiveness in detecting Reentrancy vulnerabilities on Ethereum, we used two publicly available datasets~\cite{turntherudder,8Smartbugs}, SliSE demonstrated an outstanding recall rate of 92.68\% and maintained the highest F1 scores compared to the existing tools. Through ablation experiments for each stage, we confirmed the critical importance of each stage in the overall process. Notably, precise path pruning during stage \uppercase\expandafter{\romannumeral1} is instrumental for efficient vulnerability detection. This path pruning process contributed to a significant increase in the F1 score, elevating it from 6.59\% to 78.65\%. Additionally, with the stage \uppercase\expandafter{\romannumeral2} of symbolic execution verification, we observed a substantial reduction in false positives, resulting in a precision improvement from 47.30\% to 72.16\%. Overall, SliSE provides a robust and efficient method to efficiently detect Reentrancy vulnerabilities in complex contracts. 

The main contributions of our work are as follows:
\begin{enumerate}
    \item We propose an approach that combines program slicing and symbolic execution to efficiently detect Reentrancy vulnerabilities within complex contracts. By pruning and verifying the reachability of critical paths, we achieve effective detection.
    \item We design the SliSE tool, which efficiently detects complex contracts Reentrancy vulnerabilities. Comparative experiments with state-of-the-art tools substantiate its effectiveness and efficiency.
    \item We have made our SliSE tool's source code and experimental dataset publicly available at \href{https://github.com/SliSE-SC/SliSE}{https://github.com/SliSE-SC/SliSE}.
\end{enumerate}

The paper is organized as follows. In Section~\ref{sec:bkg&mot}, we provide essential background and highlight challenges in detecting complex contract Reentrancy vulnerabilities through motivating examples. Section~\ref{sec:methodology} outlines the workflow and technical details of SliSE. We evaluate the performance and efficiency of SliSE in Section~\ref{sec:evaluation}, while Section~\ref{sec:discussion} discusses existing tool capabilities in complex contract Reentrancy vulnerability detection and threats analysis. Section~\ref{sec:relatedwork} summarizes related work, and Section~\ref{sec:conclusion} concludes this paper.

\section{background and motivation}
\label{sec:bkg&mot}
\subsection{Reentrancy Vulnerability Detection}
C-E-I (\textit{Check->Effect->Interaction}) is a critical security programming pattern in smart contracts. It requires that external interactions should only occur after all preconditions are checked and internal state updates are effected~\cite{CEI2,CEI1}. For example, when transferring tokens to an external contract (\textit{\textbf{Interaction phase}}), it is essential to perform user balance checks (\textit{\textbf{Check phase}}) and update balances (\textbf{\textit{Effect phase}}) first. In case of an attacker attempting Reentrancy, the transaction will revert due to the balance check failure in \textit{\textbf{Check phase}}, effectively preventing Reentrancy. Evaluating whether token transfers in the contract adhere to the secure C-E-I (\textit{Check->Effect->Interaction}) pattern is pivotal in Reentrancy vulnerability detection.

Reentrancy vulnerabilities typically stem from the violation of the secure \textit{Check->Effect->Interaction} (C-E-I) pattern, with attackers exploiting the smart contract's \textit{fallback mechanism}. The \textit{Check->Effect->Interaction } (C-E-I) security pattern necessitates that state changes occur before contract interactions, ensuring timely state updates. Smart contracts' \textit{fallback mechanism} activates automatically upon receiving Ether (native tokens). Some token standards, such as ERC-777~\cite{ERC777} and ERC-1155~\cite{ERC1155}, emulate the native token's \textit{fallback mechanism} using \textit{hook functions}. If the state is not promptly updated, transferring tokens to the attacker can trigger their fallback function, enabling them to hijack the logic and exploit Reentrancy vulnerabilities. This disrupts the atomicity of the transaction, leading to multiple executions from a single function call. 

Figure~\ref{MExample1} displays a Reentrancy vulnerability. The \textit{withdraw} function allows users to transfer a amount of tokens. In L5, it first verifies whether the user's balance (\textit{userBalance}) exceeds the transferred amount (\textit{\textbf{Check phase}}). If the condition is met, execution proceeds, otherwise, the transaction is reverted. L6 facilitates the transfer of \textit{$\_amount$} tokens to \textit{msg.sender} (\textbf{\textit{Interaction phase}}). Following this, L7 updates the balance of \textit{msg.sender} (\textbf{\textit{Effect Phase}}). This sequence in the \textit{withdraw} function violates the secure \textit{Check->Effect->Interaction} (C-E-I) transfer pattern, which undermines transaction atomicity and triggers Reentrancy attacks. This vulnerability arises in L6, where a transfer is initiated by \textit{msg.sender} (external user) using \textit{call.value}. Since the deduction of the attacker's balance is deferred to L7, the balance check (\textit{\textbf{Check phase}}) in L5 remains satisfied, facilitating further token transfers to the attacker. Exploiting this, an attacker can repeatedly trigger the \textit{withdraw} function through the \textit{fallback mechanism}. To rectify this, the transfer logic should follow the \textit{Check->Effect->Interaction} (C-E-I) patten, with the execution sequence as \textit{L5->L7->L6}. The balance update (L7) must precede the transfer (L6). This arrangement ensures that even if the attacker reenters the \textit{withdraw} function via the \textit{fallback mechanism}, the reentry fails due to an unsatisfied balance check condition (L5).

\begin{figure}[htb]
	\setlength{\abovecaptionskip}{0.cm}
	\begin{lstlisting}[xleftmargin=0.5cm,xrightmargin=0.5cm]
  contract ContractA {
    ...
    function withdraw(address _contractB, uint _amount) public {
        userBalance = _contractB.getBalance(msg.sender);
        require(userBalance >= _amount);
        msg.sender.call{value:_amount}("");
        _contractB.reduceBalance(msg.sender,_amount);
    }
  }
  contract ContractB {
    mapping(address => uint) balances;
    ...
    function getBalance(address _address) public view returns (uint) {
        return balances[_address];
    }
    
    function reduceBalance(address _address, uint amount) public {
        balances[_address] = balances[_address] - amount;
    }
  }
	\end{lstlisting}
	\caption{The example of Reentrancy}
	\label{MExample1}
\end{figure}

Compared to native token (ETH), ERC tokens utilize \textit{hook functions} to simulate the native token's \textit{fallback mechanism}. According to different implementation methods of \textit{fallback mechanism}, Reentrancy vulnerabilities are divided into the following two categories: 
\begin{itemize}
    \item \textbf{Reentrancy with ETH:} Transferring ETH, using the native \textit{fallback mechanism} to implement callback; 
    \item \textbf{Reentrancy with ERC Token:} Transferring ERC tokens (derivative tokens), the \textit{hook function} is used to implement callback.
\end{itemize}

Existing \textit{hook function} examples, such as ERC-777 and ERC-1155, are designed to address the problem of asset lock-in at receiving addresses. ERC-777 tokens require invoking the \textit{ERC777TokensSender} (a hook function) before state updates~\cite{ERC777}. If these updates are not synchronized, transaction atomicity can be compromised, resulting in Reentrancy vulnerabilities. Figure~\ref{MExample2} illustrates a cross-contract Reentrancy incident on Cream Finance, where the attack occurred on August 30, 2021. The attacker exploited the ERC-777 token's \textit{hook function} to borrow digital currencies twice, even though they had pledged assets only once. In the critical execution logic of the \textit{borrowFresh} function, calls \textit{borrowAllowed} to check user status (\textit{\textbf{Check phase}}) in L11, \textit{ERCToken(borrower).transfer(...)} executes the REC-777 token transfer (\textit{\textbf{Interaction phase}}) in L13, and updates the user's balance (\textit{\textbf{Effect phase}}) in L15. This violation of the \textit{Check->Effect->Interaction} (C-E-I) transfer pattern resulted in Cream Finance losing approximately \$18.8 million worth of digital assets. Notably, even though the \textit{borrowInternal} function in L5 had a Reentrancy lock (\textit{nonReentrant}), it still fell prey to a Reentrancy attack. \textbf{This underscores that a single function's Reentrancy lock cannot prevent cross-contract Reentrancy attacks.} The execution of \textit{borrow} function in L2 involves multiple internal function calls, each with distinct purposes. It is essential to extract critical semantic features to ascertain whether the contract abides by the secure \textit{Check->Effect->Interaction} (C-E-I) pattern. This emphasizes the significance of compositional security analysis when identifying Reentrancy vulnerabilities in complex contracts. Beyond examining the Reentrancy of individual functions, it is crucial to consider insecure implementations within the contract semantics.

  \begin{figure}[htb]
	\setlength{\abovecaptionskip}{0.cm}
 	\begin{lstlisting}[xleftmargin=0.5cm,xrightmargin=0.5cm]
  contract CreamFinance_Reentrancy{
    function borrow(uint borrowAmount) returns (uint) {
        return borrowInternal(borrowAmount);
    }
    function borrowInternal(uint borrowAmount) internal nonReentrant returns (uint) {
        uint error = accrueInterest();
        ...
        return borrowFresh(msg.sender, borrowAmount);
    }
    function borrowFresh(address payable borrower, uint borrowAmount) internal returns (uint) {
        uint allowed = comptroller.borrowAllowed(address(this), borrower, borrowAmount);
        ...
        ERCToken(borrower).transfer(borrowAmount);
        ...
        comptroller.borrowVerify(address(this), borrower, borrowAmount);
        return uint(Error.NO_ERROR);
    }
  }
	\end{lstlisting}
	\caption{The \textit{borrow} function causing Cream Finance's Reentrancy}
	\label{MExample2}
\end{figure}

\subsection{Motivation Examples and Challenges}

\subsubsection{Semantic Modeling of Complex Contracts} 
\begin{figure}[htb]
    \centering
    \includegraphics[width=0.8\linewidth,height=52mm]{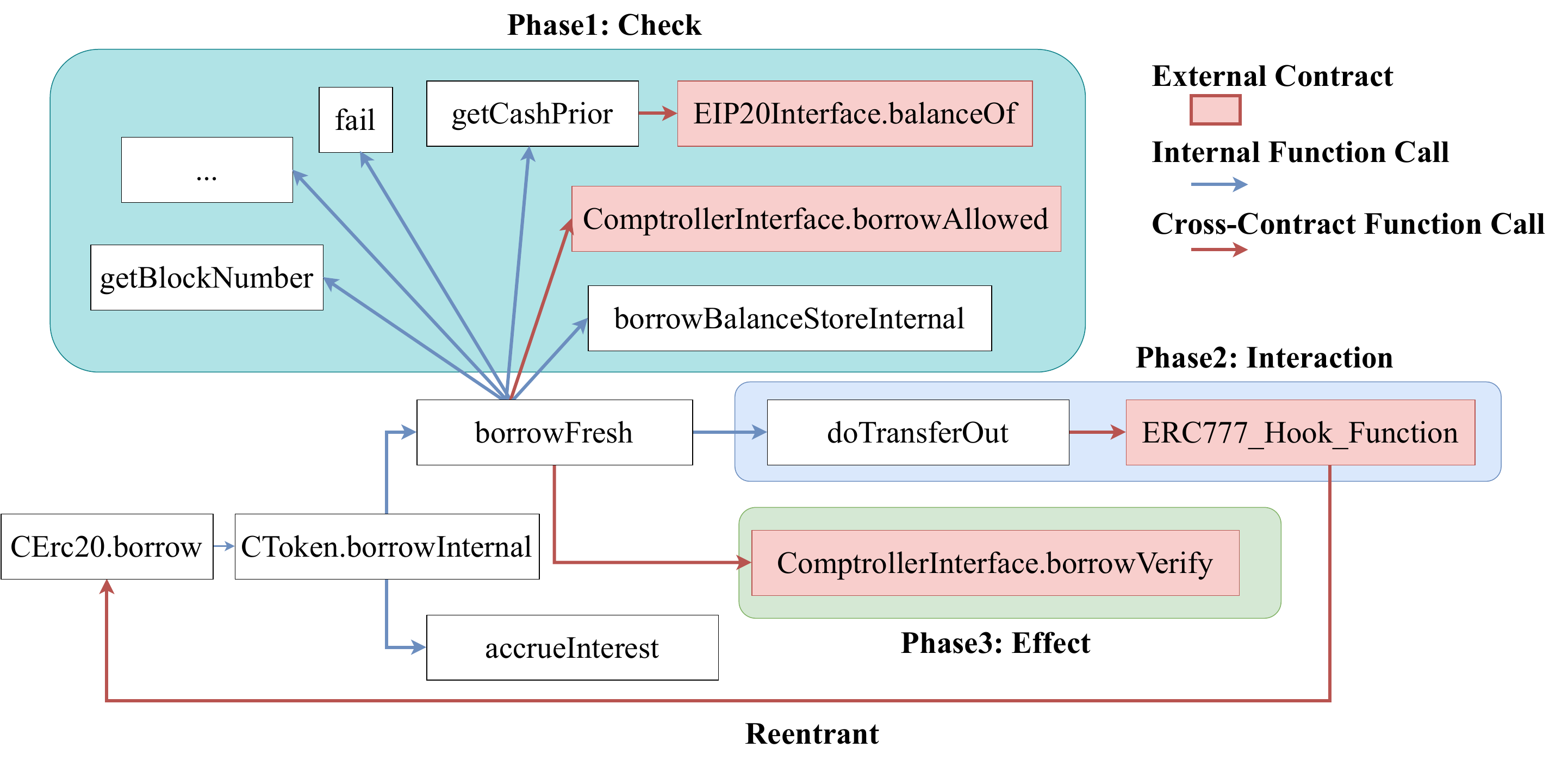}
    \caption{Execution phases and function calls of the \textit{borrow} function}
    \label{fig:motv1}
\end{figure}

The complexity of function call relationships complicates semantic analysis. Figure~\ref{fig:motv1} depicts the key function calls during the execution of the \textit{borrow} function in Figure~\ref{MExample2}. This figure highlights the intricate logic embedded within this function's implementation. Due to the modular design of smart contract functions, there's an inherent complexity in function calls. Through semantic analysis, we discern that the overall token transfer process violates the \textit{Check->Effect->Interaction} (C-E-I) pattern. Crucially, the \textbf{\textit{Interaction phase}} is executed before the \textbf{\textit{Effect phase}}, and this sequence is a pivotal factor for triggering Reentrancy. When the state is not updated promptly, the condition checked in the \textbf{\textit{Check phase}} becomes invalid. Simultaneously, external contract callbacks (attacker) intercept the execution, introducing malicious code for Reentrancy. For example, in Cream Finance's case, token transfers using the ERC-777 standard allow the receivers (attacker) to execute specific logic from the hook function upon receiving ERC-777 tokens, which is an important reason for Reentrancy attacks.

\noindent{\bf Motivation:} 
Existing approaches usually lack comprehensive compositional analysis of smart contract program semantics, typically narrowing their focus to individual functions. This limitation may result in missed vulnerabilities and false positives in detection results. As shown in Figure~\ref{fig:motv1}, the execution of the \textit{borrow} function involves complex cross-contract interactions and function calls. Tools such as Sailfish~\cite{bose2022sailfish} and Mythril~\cite{9mythril} struggle to analyze data flow during these cross-contract interactions, leading to missed vulnerability detection. On the other hand, static analysis tools like Slither~\cite{feist2019slither} primarily ensure individual function security, neglecting compositional security analysis. This limitation stems from their inability to systematically analyze state interdependencies among multiple functions, leading to a significant number of false positives in the detection results.

\noindent{\bf Challenge:}
The primary challenge in semantic modeling lies in comprehensively analyzing program dependencies and precisely capturing data dependencies during contract interactions. 

\vspace{-0.2cm}

\subsubsection{CFG Path Recovery}
\begin{figure}[H]
    \centering
    \includegraphics[width=0.45\linewidth,height=42mm]{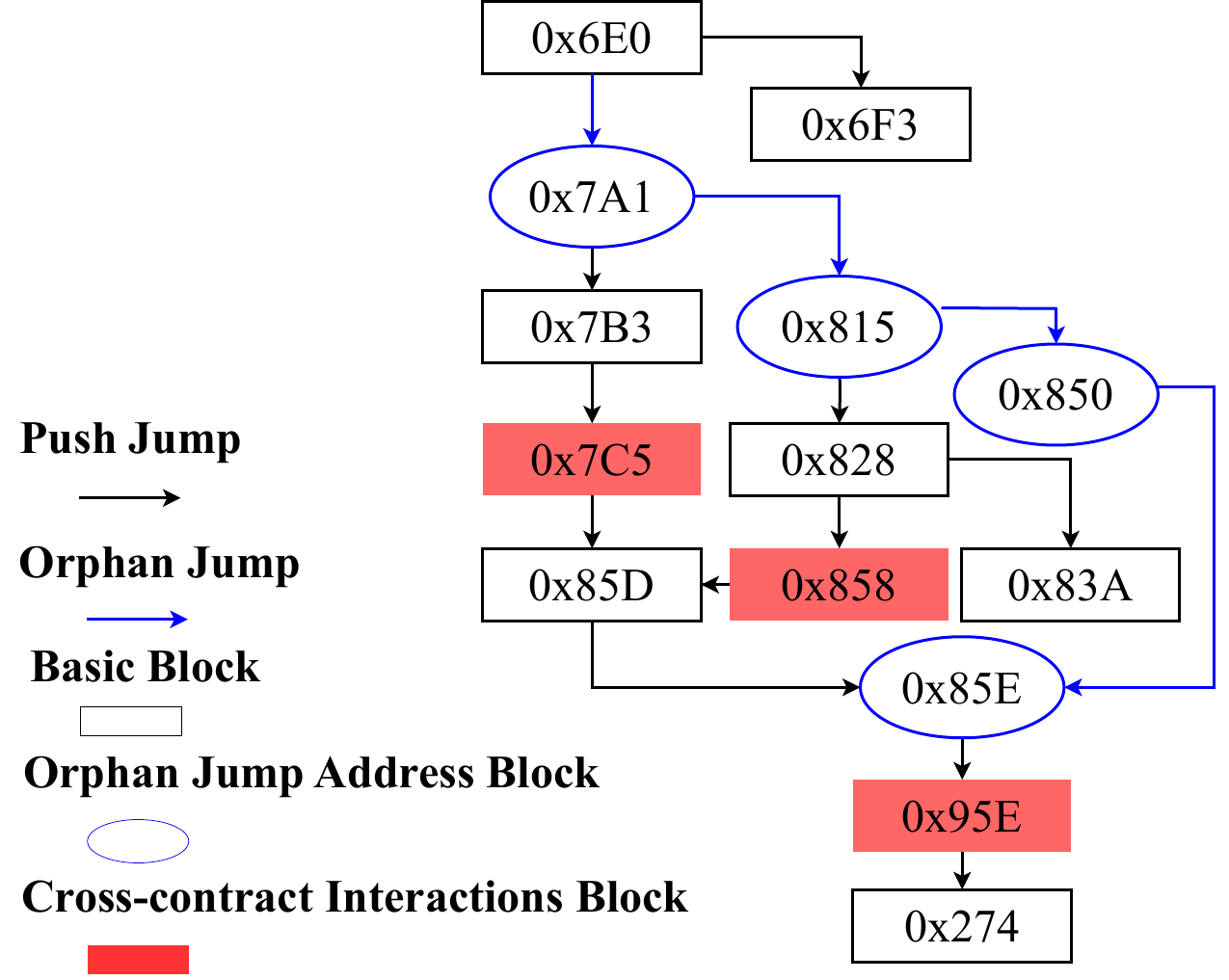}
    \caption{Control Flow Graph (CFG) of the \textit{borrow} function.}
    \label{fig:motv2}
\end{figure}

Accurately collecting and propagating path constraints are crucial for achieving effective cross-contract symbolic execution, directly impacting vulnerability detection precision. In Figure~\ref{fig:motv2}, the essential CFG paths of the \textit{borrow} function in Figure~\ref{MExample2} are depicted. The blue arrow represents the dynamic jump edge (\textit{Orphan Jump}\footnote{\textit{Orphan Jump} lacks a preceding \textit{PUSH} opcode, making its jump target address challenging to compute immediately.})~\cite{contro2021ethersolve}, where the jump address of the \textit{JUMP} opcode cannot be determined through static analysis. The blue oval block represents \textit{Orphan Jump} address block, while red filled squares represent blocks containing cross-contract interaction. As shown in Figure~\ref{fig:motv2}, the execution process of the \textit{borrow} function involves numerous dynamic jump edge calculations and cross-contract interactions. Existing tools such as Mythril~\cite{9mythril}, Sailfish~\cite{bose2022sailfish}, and Manticore~\cite{29mossberg2019manticore} encounter challenges due to difficulties in computing dynamic jump addresses and a lack of support for cross-contract analysis. Consequently, their CFG paths are incomplete, resulting in issues such as missing path constraints and incomplete path traversal. This leads to a significant number of false negatives in the detection results.

\vspace{-0.2cm}
 
\noindent{\bf Motivation:} 
Recovering dynamic jump edges is crucial for collecting path constraints and cross-contract analysis. Although traditional symbolic execution proves beneficial in CFG path recovery, it is still challenging in dynamic jump edge calculations. This limitation hinders their capability to detect vulnerabilities in complex contracts. While some tools employ static stack emulation for CFG recovery, they struggle with dynamic jump edges, leaving contract CFG paths incomplete. Furthermore, effective cross-contract analysis on complete CFG paths is essential for efficient symbolic execution. During cross-contract interactions, context switching ensures smooth collection and propagation of path constraints. However, state-of-the-art symbolic execution tools like Sailfish~\cite{bose2022sailfish} and Mythril~\cite{9mythril} frequently face difficulties in cross-contract interactions, primarily due to path losses, leading to timeouts and exacerbated path explosion.

\noindent{\bf Challenge:}
Efficiently determining dynamic jump addresses and recovering the CFG path present significant challenges.

\section{Methodology}
\label{sec:methodology}
In this section, we introduce the workflow and delve into the technical details of SliSE.

\subsection{Overview}
\begin{figure}[htb]
	\setlength{\abovecaptionskip}{0.1cm}
	\centering
	\includegraphics[width=0.60\linewidth,height=40mm]{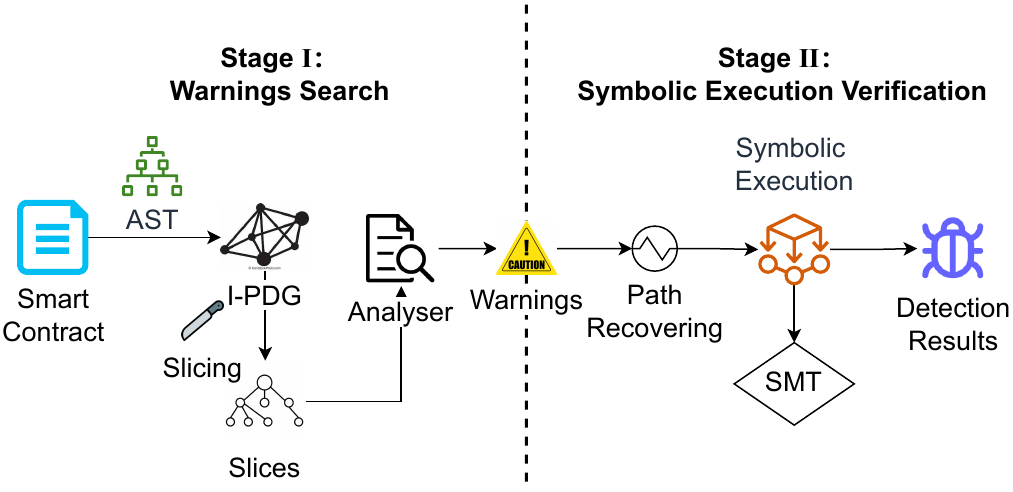}
	\caption{The workflow of SliSE} 
	\label{fig:workflow}
\end{figure}

The SliSE method efficiently detects complex contract Reentrancy vulnerabilities by combining program slicing with symbolic execution verification. The process, as illustrated in Figure~\ref{fig:workflow}, takes source code as input, reports the presence of vulnerabilities and their corresponding locations. It consists of two main stages: \textit{Warning Search} and \textit{Symbolic Execution Verification}. In Stage \uppercase\expandafter{\romannumeral1}, SliSE analyzes program dependencies through the Abstract Syntax Tree (AST) to construct the Inter-contract Program Dependency Graph (I-PDG) of the contract. It then performs slicing analysis based on Reentrancy vulnerability characteristics to identify suspicious vulnerabilities' functions and locations, which are subsequently reported as warning information. In Stage \uppercase\expandafter{\romannumeral2}, critical paths containing warning information are extracted, followed by symbolic execution to validate the reachability of these paths. This combined static analysis and symbolic execution to effectively detect Reentrancy vulnerabilities in complex contracts.

\subsection{Stage \uppercase\expandafter{\romannumeral1}: Warnings Search}
To improve search efficiency, SliSE performs program slicing analysis on the global program dependency of contracts, pruning irrelevant paths to enhance detection efficiency. Initially, SliSE compiles the contract source code to obtain the corresponding Abstract Syntax Tree (AST) information. By analyzing the AST, it constructs the \textit{Inter-contract Program Dependency Graph} (I-PDG) for the contract. Combined with the characteristics of Reentrancy vulnerabilities, program slicing is then applied to the I-PDG. This process involves verifying whether the corresponding code semantics adhere to the secure \textit{Check->Effect->Interaction} (C-E-I) pattern. The results of this analysis yield warning information about Reentrancy vulnerabilities, which serves as input for Stage \uppercase\expandafter{\romannumeral2}.
\vspace{-0.15cm}
\begin{algorithm}[h]
    \caption{Constructing I-PDG}\label{algo:construct_ipdg}
    \KwIn{Inter-contract Control Flow Graph (I-CFG)}
    \KwOut{Inter-contract Program Dependency Graph (I-PDG)}
    \SetKwFunction{ConstructIPDG}{Constructing\_I-PDG}
    \SetKwProg{Fn}{Function}{:}{}
    \Fn{\ConstructIPDG{I-CFG}}{
        I-PDG $\gets$ Initialize an empty graph\;
        \ForEach{node \textbf{in} I-CFG}{
            I-PDG.add\_node(node)\;
            \ForEach{successor \textbf{in} node.successors}{
                I-PDG.add\_node(successor)\;
                I-PDG.add\_control\_edge(node, successor)\;
                \If{modify(node, successor.variable)}{
                    I-PDG.add\_data\_edge(node, successor)\;
                }
            }
        }
        \Return I-PDG\;
    }
\end{algorithm}
\vspace{-0.15cm}

\subsubsection{Constructing Inter-contract Program Dependency Graph}
For a comprehensive analysis of smart contract program semantics, we construct the \textit{Inter-contract Program Dependency Graph} (I-PDG) by analyzing the overall dependency relationships within the contract's AST. While the existing \textit{Inter-contract Control Flow Graph} (I-CFG)~\cite{14ma2021pluto} transforms cross-contract calls into global jumps between statement blocks, offering an overview of the global control flow, it faces challenges in achieving a comprehensive global data flow analysis due to intricate execution paths and frequent cross-contract interactions. Building upon the contract's I-CFG, SliSE performs program dependency analysis between each statement block through the AST to create the \textit{Inter-Contract Program Dependency Graph} (I-PDG). In this I-PDG, nodes represent fundamental statements, and edges signify program dependency relationships involving both control and data dependencies. Algorithm~\ref{algo:construct_ipdg} outlines the construction process of the I-PDG. In this algorithm, L3--L4 iterate through each node in the I-CFG and add it to the I-PDG, while L5--L11 add the succeeding node of each to the I-PDG. Simultaneously, we analyze relationships between data definitions and usage, as well as control relationships between nodes, introducing data dependency edges and control dependency edges.

\begin{figure}[H]
    \centering
    \includegraphics[width=0.8\linewidth]{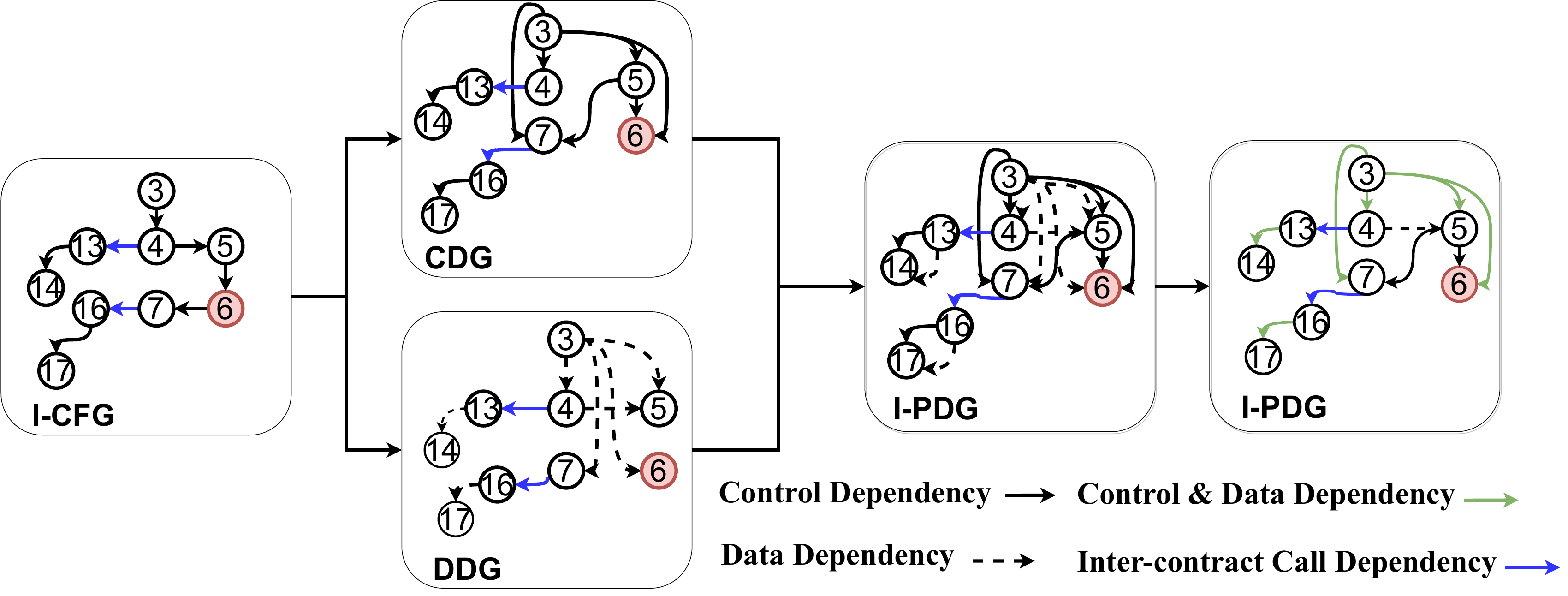}
    \caption{I-PDG construction for example in Figure~\ref{MExample1}}
    \label{fig:simple_ipdg}
\end{figure}

Figure~\ref{fig:simple_ipdg} outlines the process of constructing the \textit{Inter-contract Program Dependency Graph} (I-PDG) for contracts in Figure~\ref{MExample1}. Utilizing the foundational nodes from the I-CFG~\cite{14ma2021pluto} as a starting point, we generate subgraphs of \textit{Control Dependency Graph} (CDG) and \textit{Data Dependency Graph} (DDG) based on the control dependency relationships and the data dependency relationships between the respective nodes. Our main contribution lies in achieving a comprehensive program dependency analysis for cross-contract scenarios. Global program dependencies offer richer information for analyzing global variables, user input, data flow during cross-contract interactions, and more. As shown in Figure~\ref{fig:simple_ipdg}, the blue solid line represents the existing \textit{Inter-contract Call Dependencies}, enabling direct data flow analysis in cross-contract scenarios. This helps us understand the impact of functions invoked across contracts on global state variables and how this influence spreads.

\subsubsection{Slicing \& Analysing}
To prune irrelevant paths, we have devised specific slicing criteria based on the unique attributes of these vulnerabilities. These criteria enable precise analysis by extracting essential code sections that exhibit Reentrancy vulnerability traits. Leveraging the global program dependency, we explore relationships among cross-contract call addresses, input variables, and data flows during cross-contract interactions. These relationships are crucial for a comprehensive Reentrancy vulnerability analysis. However, since these dependency relationships are not exclusive to smart contracts, uninformed analysis could lead to significant false positives. To address this, we propose vulnerability slicing rules aligned with Reentrancy attack characteristics. These rules streamline the analysis of code relevant to Reentrancy vulnerabilities, enhancing detection efficiency. Furthermore, by considering the characteristics of Reentrancy vulnerabilities and the \textit{fallback mechanism}, we establish slicing criteria specific to two types: ETH and ERC tokens. These criteria efficiently target and analyze code relevant to Reentrancy vulnerabilities.

\begin{itemize}
    \item \textbf{Rule for Reentrancy with ETH:} \\
    Backward\_Slicing[User\_Input\_Address.call.value()]
    \item \textbf{Rule for Reentrancy with ERC Token:} \\
    Backward\_Slicing[ERC(User\_Input\_Address).call\_function()]
\end{itemize}

\textit{Rule for Reentrancy with ETH} used for vulnerabilities originating from ETH transfers, primarily utilizing the \textit{call.value()}. Consequently, our focus is on the dynamic account address (user input used as address) triggering the \textit{call.value()} function. Employing backward slicing on the I-PDG, we isolate nodes with dependencies, concentrating on sections pertinent to the Reentrancy vulnerability.


\textit{Rule for Reentrancy with ERC Token} focuses on Reentrancy originating from ERC token transfers. It centers on the function call triggered by the dynamic address contract (user input used as contract address), serving as the entry point. By conducting backward slicing of the I-PDG, we retain nodes with dependencies in the slice. For example, in Figure~\ref{MExample2}, L13 serves as the slicing entry point. This line calls the function from the dynamic address contract (ERCToken(borrower)). Given the uncertain logic of the external contract, attackers can exploit it by injecting malicious code to achieve Reentrancy. 

\begin{figure}[H]
    \centering
    \includegraphics[width=0.75\linewidth,height=28mm]{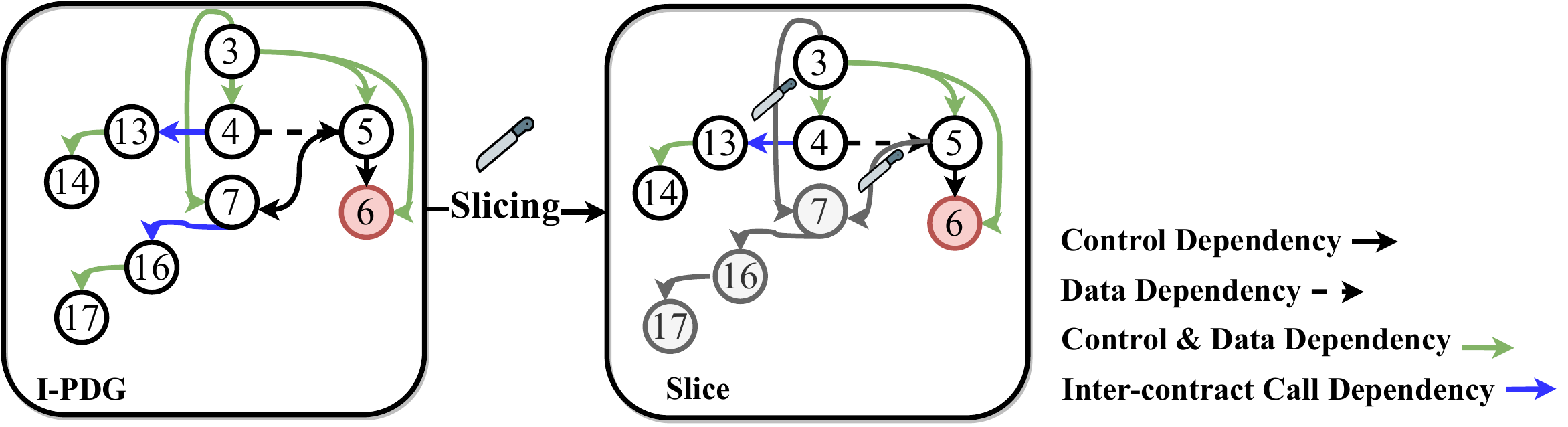}
    \caption{Slicing process for example in Figure~\ref{MExample1}}
    \label{fig:slicing}
\end{figure}

As depicted in Figure~\ref{fig:slicing}, this demonstration follows the \textit{Rule for Reentrancy with ETH} standard for the code in Figure~\ref{MExample1}. In this example, \textit{msg.sender.call.value} in L6 meets the criteria for slicing, serving as the entry point for conducting a backward slice. The resulting sliced code snippet is displayed on the right side. L3, L4, and L5 contain the code within the backward slice, while L13 and L14 are retained in the slice due to their inter-contract call dependencies. This retention preserves the integrity of dependency relationships. Our Reentrancy vulnerability program slicing criterion facilitates focused analysis of critical code segments while preserving relevant dependency relationships. This approach enables efficient and precise code analysis, mitigating the impact of unrelated code on vulnerability detection's effectiveness and accuracy.

To model and analyze semantics, we scrutinize the program dependencies of critical instruction-related statements. For instance, in Figure~\ref{MExample1}, we can extract the \textit{\textbf{Check phase}} within the code using conditional checks like \textit{require} and \textit{assert}, along with control dependency analysis. The presence of cross-contract function calls indicates the \textit{\textbf{Interaction phase}}, while statements related to variable updates in the \textit{\textbf{Check phase}} correspond to the \textit{\textbf{Effect phase}}. SliSE employs this information to slice the contract's I-PDG, assessing its adherence to the secure development pattern \textit{Check->Effect->Interaction} (C-E-I), and generating pertinent warning messages.

\begin{figure}[H]
    \centering
    \includegraphics[width=0.75\linewidth,height=56mm]{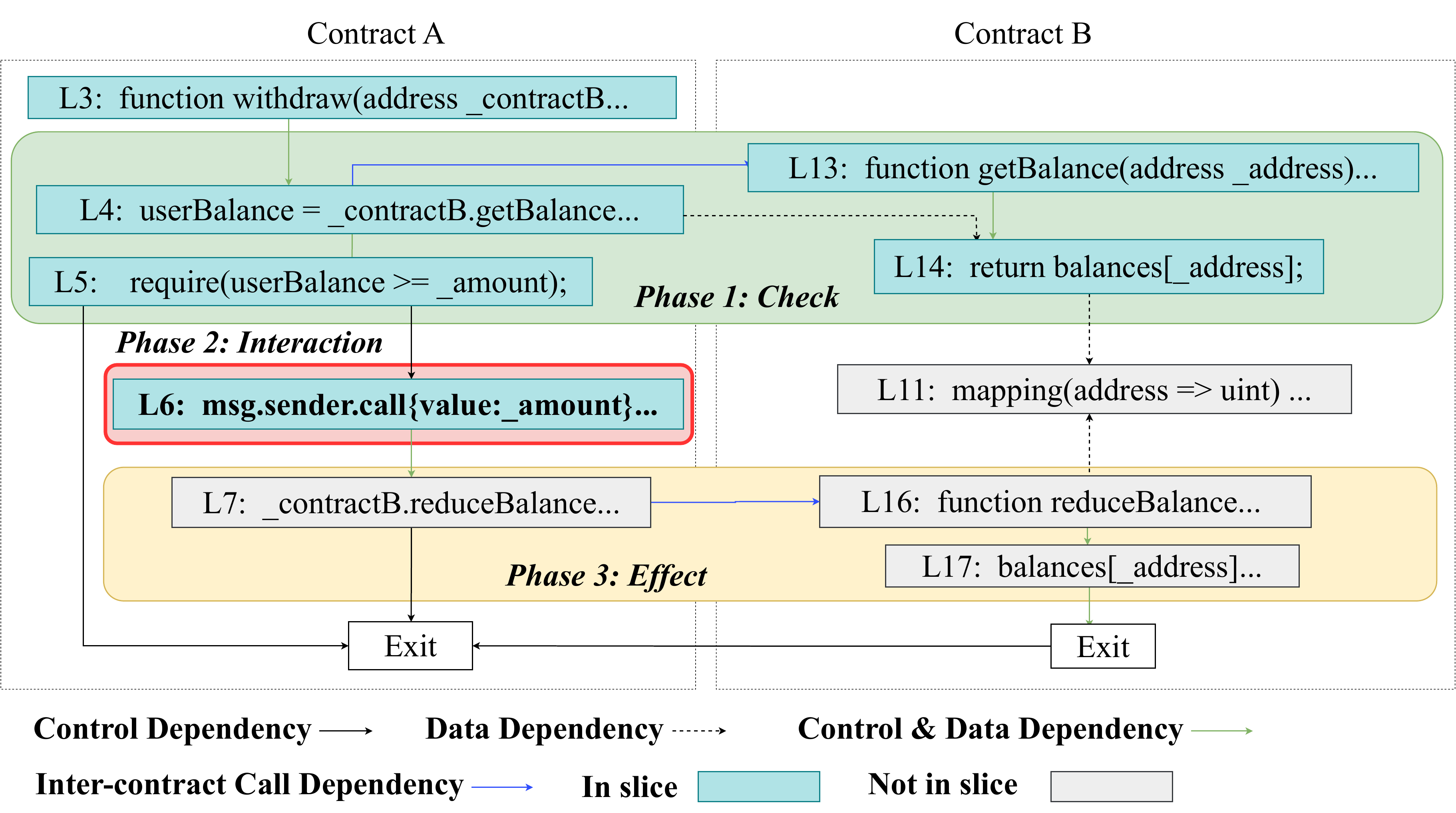}
    \caption{Analysis of Program Slicing for Figure~\ref{MExample1}}
    \label{fig:slice_example}
\end{figure}

Figure~\ref{fig:slice_example} illustrates the analysis procedure of the slice corresponding to the code presented in Figure~\ref{MExample1}. Within the slice code, the \textit{require()} in L5 represents the \textit{\textbf{Check phase}}, involving the \textit{balances[user]} variable. Additionally, \textit{msg.sender.call.value()} in L6 signifies the \textbf{\textit{Interaction phase}}, aligning with the slicing rule of the \textit{Rule for Reentrancy with ETH}. It is noteworthy that no actions are taken to alter the \textit{balances[user]} variable (modification occurs in L7), indicating that the \textit{\textbf{Effect phase}} follows the \textit{\textbf{Interaction phase}}. This observation plays a crucial role in identifying the Reentrancy vulnerability. When an attacker reenters through the external contract, the state of the \textit{\textbf{Check phase}} remains unchanged (\textit{balances[user]} does not decrease). Consequently, it can be identified as a Reentrancy vulnerability. We compile the location details of the vulnerable function to generate warning information.
\vspace{-0.1cm}
\vspace{-0.2cm}

\subsection{Stage \uppercase\expandafter{\romannumeral2}: Symbolic Execution Verification}

To reduce false positives and ensure reliable vulnerability detection, SliSE employs symbolic execution for path reachability verification. Initially, SliSE addresses the challenge of calculating dynamic jump addresses (\textit{Orphan Jump}) by using Algorithm~\ref{algo:recovery_cfg} combined with SSA\footnote{SSA is a property of intermediate languages that mandates each variable to be assigned only once, enabling features like constant propagation analysis.} (Static Single Assignment) to ensure complete CFG paths. Based on the warning information from Stage \uppercase\expandafter{\romannumeral1}, critical paths are traversed, and path constraints are collected. These constraints are stored in the \textit{Symbolic Register} for later access and retrieval. They are then validated using the \textit{Z3-solver} to confirm path reachability and, consequently, the existence of vulnerabilities.

\subsubsection{Path Recovering}
To efficiently recover CFG paths, we utilize \textit{constant propagation analysis} combined with SSA to determine the target addresses of \textit{Orphan Jumps}. Initially, the bytecode is divided into multiple blocks, and some blocks connection are recovered through static stack emulation. However, static stack emulation can only recover jump edges for \textit{Push Jumps}\footnote{\textit{Push Jump} is immediately preceded by a \textit{PUSH} opcode, making its jump target address easily calculable.}, and it cannot ascertain the target of an \textit{Orphan Jump} within the current block. This limitation results in the inability to recover the edges of \textit{Orphan Jumps}, leaving the CFG incomplete.

\begin{figure}[H]
    \centering
    \includegraphics[width=0.45\linewidth,height=58mm]{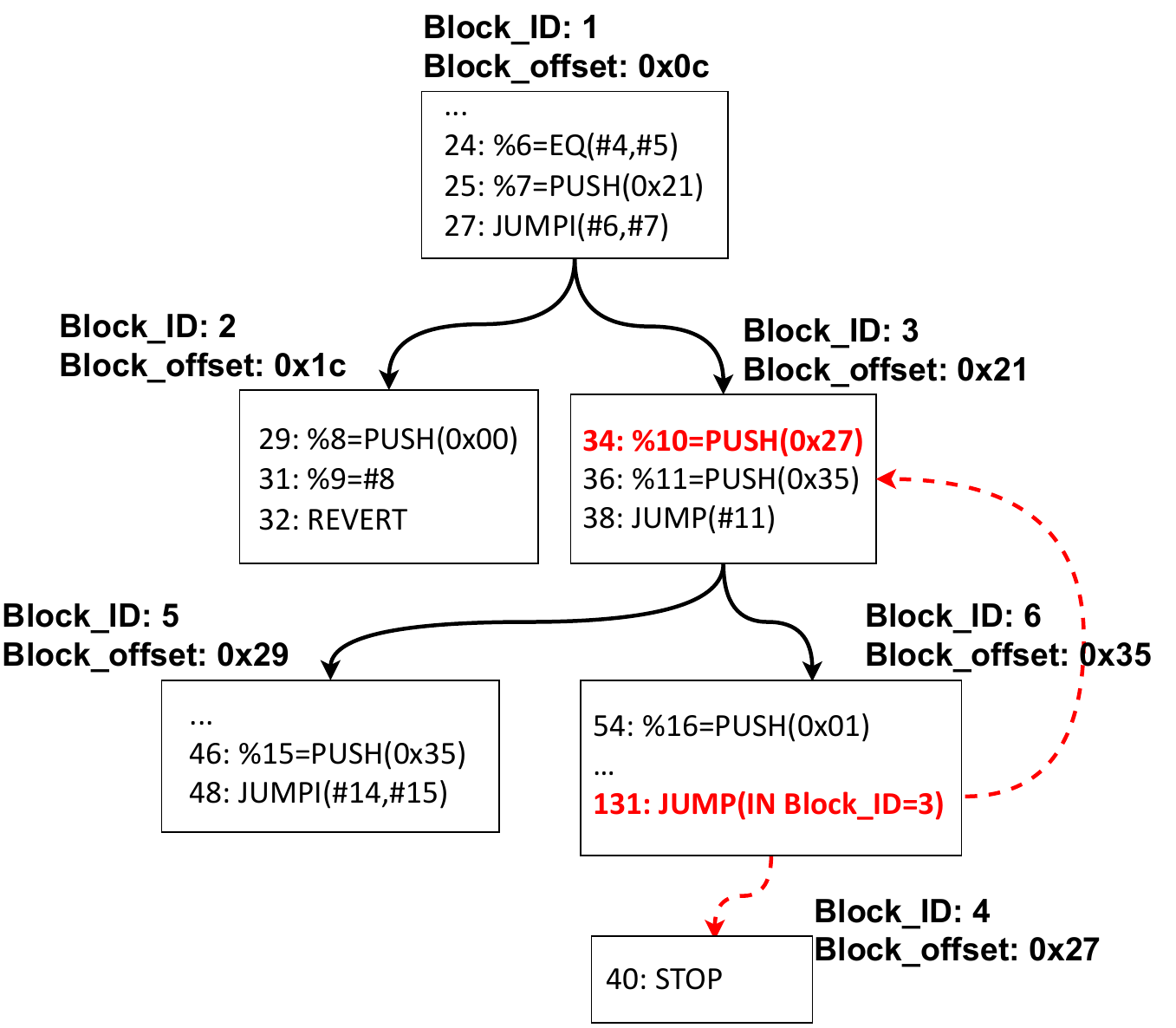}
    \caption{Recovering the CFG with SSA}
    \label{fig:ssa_cfg}
\end{figure}

Figure~\ref{fig:ssa_cfg} illustrates an example of recovering the CFG with SSA. In the incomplete CFG obtained through static stack emulation, a \textit{JUMP} instruction exists within Block\_ID 6, but this block lacks successor blocks, creating what is known as an \textit{Orphan Jump}. The inherent limitation of static stack emulation in addressing the \textit{Orphan Jump} issue lies in its incomplete analysis of variable transitions within the current block. It cannot effectively analyze the flow of values along the execution path of the current block, making it incapable of solving the \textit{Orphan Jump} problem effectively. By leveraging the property of SSA, which ensures that each variable is assigned only once, we can perform global value propagation analysis along the execution path of the \textit{Jump} block. The SSA representation clearly highlights the relationships between variable definitions and their uses within each block. If, prior to the \textit{Jump} instruction, there exists a variable that is defined but not used, that variable represents the \textit{stacktop value} and serves as the target of the \textit{Orphan Jump}. If no \textit{stacktop value} is found in the current block, the search continues in the preceding blocks. For instance, in Block\_ID 6, where a \textit{JUMP} instruction is present, the \textit{stacktop value} can be located in its predecessor block (Block\_ID 3). This approach ensures the precise recovery of target addresses for \textit{Orphan Jumps}, obtaining a complete CFG.
\vspace{-0.1cm}
\vspace{-0.2cm}

To search the target addresses, we utilize constant propagation analysis with SSA to calculate the dynamic target addresses of \textit{Orphan Jumps}. Algorithm~\ref{algo:recovery_cfg} outlines the process for CFG path recovery. This process involves transforming the bytecode within each block into SSA form, ensuring that each numeric value is assigned only once. As we traverse the blocks of the incomplete CFG, we update the SSA global variables (\textit{ssaVariables}) based on operations that involve variable assignment or access (e.g., \textit{PUSH, DUP, POP, SWAP})~\cite{wood2014ethereum}. For example, in L8--L9, we identify the jump target by searching for unused variables in the current block. Unused variables refer to those defined in the block but are not utilized. If such a variable exists within the current block, it becomes the jump target. However, if it does not, we perform a search within the preceding blocks along the CFG path where the current block is located. The \textit{findUnusedVar} function facilitates this iterative search process. This algorithm ensures accurate propagation of values along the execution path, facilitating the precise recovery of target addresses for \textit{Orphan Jumps} embedded within the CFG.

 \subsubsection{Symbolically Verifying Path Feasibility}
To ensure the accuracy of Reentrancy vulnerability detection, we perform a reachability analysis of the warning path information. While Stage \uppercase\expandafter{\romannumeral1} conducts program slicing analysis to generate warning information related to violating the \textit{Check->Effect->Interaction} (C-E-I) pattern, it lacks verification of the reachability of the warning paths. Specifically, contract \textit{Path Protective Techniques} (PPTs)~\cite{11ye2020clairvoyance}, such as mutex locks and permission checks, can limit the occurrence of Reentrancy vulnerabilities. Static analysis tools often struggle to accurately identify these protective techniques, leading to numerous false positives. To address this, SliSE collects warning path constraints and validates their reachability. When it encounters assignments or accesses involving symbolic values, it uses the \textit{symbolic register} for symbolic state manipulation. Considering the conditions for Reentrancy vulnerability and the characteristics of smart contracts, it collects constraints and sends them to \textit{Z3-solver} for constraint calculation to prove path reachability, significantly enhancing the accuracy of vulnerability detection.

To ensure the integrity of path constraints, we manage symbolic expressions in a key-value pair format for further analysis. By examining relevant operations related to state storage and utilizing Z3 library functions, we store symbolic expressions as key-value pairs in the \textit{Symbolic Register}. The access and storage of symbolic states are crucial for collecting path constraints. By querying the \textit{Symbolic Register}, SliSE determines whether the contract address involved in external contract function calls is a symbolic value, assessing whether the logic of the called contract is controlled by external users. This also facilitates quick assessment of whether a contract might be susceptible to hijacking by malicious callback functions, streamlining the efficient propagation analysis of cross-contract path constraints in various contract contexts during symbolic execution verification.

\vspace{-0.1cm}
\begin{algorithm}[htb]
    \caption{Recovery CFG}
    \label{algo:recovery_cfg}
    
    \SetKwFunction{Update}{Update}
    \SetKwFunction{getUnusedVariables}{getUnusedVariables}
    \SetKwFunction{findUnusedVar}{findUnusedVar}
    \SetKwFunction{RecoveryCFG}{RecoveryCFG}
    
    \KwIn{CFG}
    \KwOut{Reconstructed CFG}
    
    \SetKwProg{Fn}{Function}{:}{}
    
    \Fn{\RecoveryCFG{CFG}}{
        \ForEach{block \textbf{in} CFG.blocks}{
            jumpTarget $\gets$ \findUnusedVar(block)\;
            CFG.add\_edge(block, blockAt(jumpTarget))\;
        }
    }
    \Fn{\findUnusedVar{block}}{
        \ForEach{preBlock \textbf{in} block.predecessors}{
            StackTopValue $\gets$ \getUnusedVariables(preBlock)\;
            \Return StackTopValue\;
        }
    }
\end{algorithm}

\vspace{-0.2cm}
\section{Evaluation}
\label{sec:evaluation}
In this section, we evaluated SliSE's performance by comparing the detection results with eight state-of-the-art tools and analyzing the impact of each stage on the overall detection process. We will address the following research questions:
\begin{enumerate}[label=\textbf{RQ\arabic*.}]
    \item How effective is SliSE in detecting complex contracts Reentrancy vulnerabilities?
    \item How effective is SliSE in detecting Reentrancy vulnerabilities on Ethereum?
    \item What is the impact of pruning in Stage \uppercase\expandafter{\romannumeral1}?
    \item What is the impact of symbolic execution verification in Stage \uppercase\expandafter{\romannumeral2}?
\end{enumerate}

\subsection{Experimental Setup}
The experiments were conducted on a computer running Ubuntu 18.04.1 LTS, equipped with a 16-core Intel(R) Xeon(R) Gold 5217 processor. We set up the experiments by downloading images or manually configuring. We used default parameters and a time budget of 300 seconds to ensure that the experiments did not excessively consume time, aligning with the approach suggested in \cite{turntherudder}. 

Tools selection following with \cite{zheng2023dappscan}, the experiment using a set of analysis tools, including Slither~\cite{feist2019slither}, Mythril~\cite{9mythril}, Securify~\cite{tsankov2018securify}, Smartian~\cite{choi2021smartian}, and Sailfish~\cite{bose2022sailfish}. Additionally, to facilitate a comparative analysis of the performance of existing symbolic execution detection tools, we also integrated three symbolic execution tools, i.e., Oyente~\cite{luu2016making}, Osiris~\cite{13torres2018osiris}, and Manticore~\cite{29mossberg2019manticore}.

The experimental datasets include DB1, DB2, and DB3. DB1 is sourced from open-source DApp auditing projects, representing off-chain versions with complex contract logic. DB2 comprises real on-chain contracts, most of which have been labeled as positive by existing tools, posing a significant detection challenge. DB3 is the simplest and most widely utilized dataset. Smart contract of these datasets all originate from real-world production environments, compared and analyzed through \textit{Precision} ($\frac{TP}{TP+FP}$), \textit{Recall} ($\frac{TP}{TP+FN}$), and \textit{F1 score} ($\frac{2*precision*recall}{precision+recall}$). DB1 was employed to answer RQ1, RQ3, and RQ4, while DB2 and DB3 were utilized to answer RQ2, RQ3, and RQ4.

\begin{enumerate}[label=\textbf{DB\arabic*.}]
    \item \textit{Complex Contract Dataset}~\cite{zheng2023dappscan}. Zheng et al. compiled a complex contract dataset that encompasses 895 vulnerabilities from 1,322 open-source DApp audit projects provided by 30 blockchain security companies, with a total of 81 positive labels for Reentrancy vulnerabilities.
    \item \textit{Ethereum Contract Dataset}~\cite{turntherudder}. Zheng et al. used existing detection tools to analyze 230,548 verified contracts from Etherscan and obtained this dataset through manual inspection. The dataset includes 21,212 contracts identified as positive Reentrancy vulnerabilities using six Reentrancy detection tools, with 41 of contracts manually verified as true positives.
    \item \textit{SmartBugs Dataset}~\cite{8Smartbugs}. The SmartBugs includes 143 contracts, of which 31 are labeled with Reentrancy vulnerabilities.
\end{enumerate}


Research indicates that manually labeling smart contracts is highly error-prone~\cite{turntherudder}, and achieving precise labeling for all Ethereum smart contracts is a formidable challenge. To ensure the accuracy of our experimental analysis, we conducted RQ2 experiments using the DB2 and DB3 datasets. These datasets have has been presented at top software engineering conferences and are widely recognized within the academic community, serving as a reliable source of ground truth for Ethereum contracts. We have compared key attributes among these datasets, as depicted in Table~\ref{tab:datasets}. Notably, the average lines of code (Loc) in DB1 are approximately 24.9 times greater than in DB3 and 3.4 times greater than in DB2. Similarly, the average number of functions is 29.9 times higher in DB3 and 3.6 times higher in DB2 compared to DB1. These findings underscore the high complexity of DB1 dataset from the real production environment, and we will delve into a quantitative analysis of DB1's complexity in Section~\ref{sec:discussion}.

\vspace{-0.1cm}
\begin{table}[htb]
    \begin{center}
        \caption{Statistics for different datasets}
        \label{tab:datasets}
        \begin{threeparttable}
            \resizebox{0.68\columnwidth}{!}{
            \large 
            \begin{tabular}{c|ccccc}
            \hline
            \textbf{Dataset}   & \textbf{Loc} & \textbf{\# of Subcontracts} & \textbf{\# of Functions} & \textbf{\# P} & \textbf{\# N} \\ \hline
            \textbf{DB1}       & 1812.5       & 12                          & 197.4                    & 81            & 814           \\
            \textbf{DB2}      & 534.5        & 6.0                         & 54.2                     & 41            & 21171         \\
            \textbf{DB3} & 72.8         & 1.5                         & 6.6                      & 31            & 112           \\ \hline
            \end{tabular}
            }
        \end{threeparttable}
    \end{center}
\end{table}
\vspace{-0.20cm}

\subsection{Effectiveness of Detecting Reentrancy for Complex Contracts}

\begin{table*}[htb]
        \fontsize{14}{19.8}\selectfont 
	\setlength\tabcolsep{3pt}
	\setlength{\abovecaptionskip}{0.1cm}
	\begin{center}
		\caption{Statistics of Detection Results by Different Tools}
		\label{tab:RQ1}
		\begin{threeparttable}
			\resizebox{\textwidth}{!}{
\begin{tabular}{c|ccc|ccc|ccc|ccc|ccc|ccc|ccc|ccc|ccc}
\hline
\textbf{Tool} & \multicolumn{3}{c|}{\textbf{Mythril}}      & \multicolumn{3}{c|}{\textbf{Securify}}     & \multicolumn{3}{c|}{\textbf{Slither}}      & \multicolumn{3}{c|}{\textbf{Oyente}}       & \multicolumn{3}{c|}{\textbf{Osiris}}       & \multicolumn{3}{c|}{\textbf{Manticore}}    & \multicolumn{3}{c|}{\textbf{Smartian}}     & \multicolumn{3}{c|}{\textbf{Sailfish}}     & \multicolumn{3}{c}{\textbf{Ours}}          \\ \hline
\textbf{Dataset}    & \textbf{DB1} & \textbf{DB2} & \textbf{DB3} & \textbf{DB1} & \textbf{DB2} & \textbf{DB3} & \textbf{DB1} & \textbf{DB2} & \textbf{DB3} & \textbf{DB1} & \textbf{DB2} & \textbf{DB3} & \textbf{DB1} & \textbf{DB2} & \textbf{DB3} & \textbf{DB1} & \textbf{DB2} & \textbf{DB3} & \textbf{DB1} & \textbf{DB2} & \textbf{DB3} & \textbf{DB1} & \textbf{DB2} & \textbf{DB3} & \textbf{DB1} & \textbf{DB2} & \textbf{DB3} \\ \hline
\textbf{\# TP}   & 5            & 8            & 13           & 0            & 31           & 29           & 8            & 36           & 30           & 0            & 32           & 28           & 0            & 34           & 29           & 0            & 0            & 0            & 0            & 7            & 19           & 0            & 26           & 25           & 70           & 38           & 30           \\
\textbf{\# FP}   & 22           & 15492        & 51           & 0            & 2356         & 44           & 140          & 18346        & 67           & 0            & 481          & 39           & 1            & 476          & 31           & 0            & 22           & 0            & 0            & 15           & 9            & 2            & 2270         & 31           & 27           & 31           & 14           \\
\textbf{\# FN}   & 76           & 33           & 18           & 81           & 10           & 2            & 73           & 5            & 1            & 81           & 9            & 3            & 81           & 7            & 2            & 81           & 41           & 31           & 81           & 34           & 12           & 81           & 15           & 6            & 11           & 3            & 1            \\
\textbf{\# TN}   & 792          & 5679         & 61           & 814          & 18815        & 68           & 676          & 2825         & 45           & 814          & 20690        & 73           & 813          & 20695        & 81           & 814          & 21149        & 112          & 814          & 21156        & 103          & 812          & 18901        & 81           & 787          & 21140        & 98           \\ \hline
\textbf{P}       & 18.52\%      & 0.05\%       & 20.31\%      & 0.00\%       & 1.30\%       & 39.73\%      & 5.41\%       & 0.20\%       & 30.93\%      & 0.00\%       & 6.24\%       & 41.79\%      & 0.00\%       & 6.67\%       & 48.33\%      & 0.00\%       & 0.00\%       & 0.00\%       & 0.00\%       & 31.82\%      & 67.86\%      & 0.00\%       & 1.13\%       & 44.64\%      & 72.16\%      & 55.07\%      & 68.18\%      \\ \hline
\textbf{R}       & 6.17\%       & 19.51\%      & 41.94\%      & 0.00\%       & 75.61\%      & 93.55\%      & 9.88\%       & 87.80\%      & 96.77\%      & 0.00\%       & 78.05\%      & 90.32\%      & 0.00\%       & 82.93\%      & 93.55\%      & 0.00\%       & 0.00\%       & 0.00\%       & 0.00\%       & 17.07\%      & 61.29\%      & 0.00\%       & 63.41\%      & 80.65\%      & 86.42\%      & 92.68\%      & 96.77\%      \\ \hline
\textbf{F}       & 9.26\%       & 0.10\%       & 27.37\%      & 0.00\%       & 2.55\%       & 55.77\%      & 6.99\%       & 0.39\%       & 46.88\%      & 0.00\%       & 11.55\%      & 57.14\%      & 0.00\%       & 12.34\%      & 63.74\%      & 0.00\%       & 0.00\%       & 0.00\%       & 0.00\%       & 22.22\%      & 64.41\%      & 0.00\%       & 2.23\%       & 57.47\%      & 78.65\%      & 69.09\%      & 80.00\%      \\ \hline
\end{tabular}
			}
                \begin{tablenotes}
                   \tiny
                     \item[] 
                     \hspace{0.5cm} * \textbf{P} represents \textit{Precision}, \textbf{R} represents \textit{Recall}, and \textbf{F} represents \textit{F1 score}.
		    \end{tablenotes}
		\end{threeparttable}
	\end{center}
\end{table*}

To evaluate SliSE's capability in detecting Reentrancy vulnerabilities within complex contracts, we conducted experiments comparing it with eight state-of-the-art tools using the DB1 dataset. The results, presented in Table~\ref{tab:RQ1}, demonstrate the detection performance of different tools. It is evident that existing tools achieve a maximum F1 score of less than 10\%, and most of them correctly detect 0 Reentrancy vulnerabilities. 
This underscores the limitations of existing tools in effectively detecting vulnerabilities in real-world, complex DApps. In contrast, SliSE perform the best, achieving an impressive F1 score of 78.65\% on DB1, a significant improvement compared to the highest F1 score of 9.26\% achieved by existing tools.

Table~\ref{tab:RQ1-time} compares the detection times of different tools. SliSE takes an average of 25.26 seconds for DB1 detection. While this is higher than static analysis tools like Slither, it falls within a moderate range compared to most symbolic execution tools.

The analysis of experimental results reveals that many existing symbolic execution tools face challenges in precisely searching paths while detecting vulnerabilities in complex contracts. This is due to their limited state exploration capabilities, resulting in incomplete path identification and the omission of numerous paths. This inefficiency slows down program execution and generates a notable number of false negatives. Therefore, precise path pruning is essential for efficient and effective symbolic execution vulnerability detection.

\vspace{-0.10cm}
\begin{table}[htb]
    \begin{center}
        \caption{Comparison of average detection times}
        \label{tab:RQ1-time}
        \begin{threeparttable}
            \resizebox{0.88\columnwidth}{!}{
            \large 
\begin{tabular}{c|ccccccccc}
\hline
\textbf{Dataset} & \textbf{Mythril} & \textbf{Securify} & \textbf{Slither} & \textbf{Oyente} & \textbf{Osiris} & \textbf{Manticore} & \textbf{Smartian} & \textbf{Sailfish} & \textbf{Ours} \\ \hline
\textbf{DB1}     & 157.17           & 26.89             & 7.11             & 8.76            & 14.15           & 189.8              & 238.79            & 7.04              & 25.26         \\
\textbf{DB2}     & 276.36           & 101.18            & 4.97             & 13.53           & 154.54          & 290.77             & 298.38            & 1.19              & 6.01          \\
\textbf{DB3}     & 230.49           & 84                & 4.37             & 9.33            & 141.84          & 248.63             & 264.11            & 5.32              & 1.94          \\ \hline
\end{tabular}
            }
        \end{threeparttable}
    \end{center}
\end{table}
\vspace{-0.20cm}

\textbf{Answer to RQ1:} SliSE outperformed other tools in detecting vulnerabilities within complex contracts, achieving an impressive F1 score of 78.65\%, significantly surpassing the maximum of 9.26\% achieved by other tools. The precise path pruning techniques employed by SliSE are instrumental in ensuring effective and efficient vulnerability detection.

\subsection{Effectiveness of Detecting Reentrancy on Ethereum Contracts}


To evaluate SliSE's effectiveness in detecting Reentrancy vulnerabilities on Ethereum contracts, we conducted experiments with DB2 and DB3. These datasets originated from previous research on smart contract security~\cite{turntherudder,8Smartbugs}, containing 68,610 contracts that were manually annotated and widely recognized as the ground truth for Reentrancy vulnerabilities on Ethereum contracts.

We conducted comparative experiments using eight existing tools, as shown in Table~\ref{tab:RQ1}. SliSE's superiority is evident, achieving detection recall rates exceeding 90\%. On DB2, it significantly outperforms existing tools, demonstrating robust detection capabilities for complex contracts. Notably, apart from Manticore, existing tools exhibit relatively average performance on DB3, with most F1 scores reaching 45\% or higher. However, when dealing with the more complex DB2, the vulnerability detection capabilities of existing tools notably decline. The majority of these tools yield F1 scores below 15\% due to a high number of false positives. Furthermore, in the time comparison for DB2 and DB3 in Table~\ref{tab:RQ1-time}, symbolic execution tools like Mythril and Manticore still struggle with path explosion issues, with average detection times exceeding 230 seconds. In contrast, SliSE's detection time is comparable to that of static analysis tools, averaging less than 7 seconds.

\vspace{-0.10cm}
\begin{figure}[H]
	\setlength{\abovecaptionskip}{0.cm}
	\begin{lstlisting}
  function withdrawAll() external {
      uint256 amount = address(this).balance;
      bool success = msg.sender.call{value: amount}("");
      require(success, "Transfer failed");
  }
	\end{lstlisting}
    \caption{False positive of Reentrancy without economic loss}
	\label{fig:RQ2_exam2}
\end{figure}
\vspace{-0.20cm}

Through analysis, we have identified that false positives (FP) are primarily associated with \textit{specific semantic designs}, which refer to program implementations tailored for specific scenarios. One example is \textit{Reentrancy can occur without causing economic losses}, meaning that Reentrancy logic exists in the contract but does not result in any losses to the victim. These situations often arise from specific program design choices that prevent Reentrancy but require contextual analysis. SliSE doesn not account for these special semantic designs, leading to false positives. In Figure~\ref{fig:RQ2_exam2}, the code satisfies the conditions for a Reentrancy vulnerability, allowing Reentrancy through the \textit{call.value()} in L3. However, at this point, the contract's balance has already been fully transferred, causing the transaction to revert due to insufficient balance during the Reentrancy. In reality, no economic loss occurs. The lack of specific semantic analysis significantly results in false positives in the detection outcomes.

\textbf{Answer to RQ2:} SliSE's advantage of detecting Reentrancy vulnerabilities in Ethereum smart contracts is evident when compared to existing tools. It consistently achieves a recall rate exceeding 90\% and maintains higher F1 scores compared to the best results achieved by existing tools.

\subsection{Impact of Pruning in Stage \uppercase\expandafter{\romannumeral1}}
In Stage \uppercase\expandafter{\romannumeral1}, SliSE performs program slicing to analyze the contract's I-PDG, pruning irrelevant code segments for targeted analysis. To evaluate the impact of pruning in Stage \uppercase\expandafter{\romannumeral1} on overall detection, we conducted comparative experiments in two modes: \textit{Stage \uppercase\expandafter{\romannumeral2}} and \textit{Stage \uppercase\expandafter{\romannumeral1} \& \uppercase\expandafter{\romannumeral2} combined} (representing without and with Stage \uppercase\expandafter{\romannumeral1}). By comparing results with and without pruning in Stage \uppercase\expandafter{\romannumeral1}, we assessed its effect, as shown in Table~\ref{tab:RQ2}. The impact of pruning is minimal when dealing with datasets of low complexity, such as DB2 and DB3. Pruning in Stage \uppercase\expandafter{\romannumeral1} results in less than a 10\% increase in the overall F1 score on DB3. However, for complex contract vulnerability detection on DB1, Stage \uppercase\expandafter{\romannumeral1} pruning significantly enhances accuracy. The F1 score improves from 6.59\% to 78.65\%, achieving a 11 times increase in results.

\begin{table*}[htb]
        \large
	\setlength\tabcolsep{3pt}
	\setlength{\abovecaptionskip}{0.1cm}
	\begin{center}
		\caption{Statistics of Detection Results by Different Tools}
		\label{tab:RQ2}
		\begin{threeparttable}
			\resizebox{0.88\textwidth}{!}{
\begin{tabular}{c|ccc|ccc|ccc}
\hline
\textbf{Dataset}   & \multicolumn{3}{c|}{\textbf{DB1}}                            & \multicolumn{3}{c|}{\textbf{DB2}}                            & \multicolumn{3}{c}{\textbf{DB3}}                             \\ \hline
\textbf{Tools}     & \textbf{Stage \uppercase\expandafter{\romannumeral1}} & \textbf{Stage \uppercase\expandafter{\romannumeral2}} & \textbf{Stage \uppercase\expandafter{\romannumeral1} \& \uppercase\expandafter{\romannumeral2}} & \textbf{Stage \uppercase\expandafter{\romannumeral1}} & \textbf{Stage \uppercase\expandafter{\romannumeral2}} & \textbf{Stage \uppercase\expandafter{\romannumeral1} \& \uppercase\expandafter{\romannumeral2}} & \textbf{Stage \uppercase\expandafter{\romannumeral1}} & \textbf{Stage \uppercase\expandafter{\romannumeral2}} & \textbf{Stage \uppercase\expandafter{\romannumeral1} \& \uppercase\expandafter{\romannumeral2}} \\ \hline
\textbf{\# TP}        & 70              & 3               & 70                       & 38              & 16              & 38                       & 30              & 25              & 30                       \\
\textbf{\# FP}        & 78              & 7               & 27                       & 146             & 11               & 31                       & 56              & 1               & 14                       \\
\textbf{\# FN}        & 11              & 78              & 11                       & 3               & 25              & 3                        & 1               & 5               & 1                        \\
\textbf{\# TN}        & 736             & 807             & 787                      & 21025           & 21160           & 21140                    & 56              & 112             & 98                      \\ \hline
\textbf{Precision} & 47.30\%         & 30.00\%        & 72.16\%                  & 20.65\%         & 59.26\%        & 55.07\%                  & 34.88\%         & 96.15\%        & 68.18\%                  \\ \hline
\textbf{Recall}    & 86.42\%         & 3.70\%          & 86.42\%                  & 92.68\%         & 39.02\%         & 92.68\%                  & 96.77\%         & 83.33\%         & 96.77\%                  \\ \hline
\textbf{F1}        & 61.14\%         & 6.59\%          & 78.65\%                  & 33.78\%         & 47.06\%         & 69.09\%                  & 51.28\%         & 89.29\%         & 80.00\%                  \\ \hline
\end{tabular}
			}
		\end{threeparttable}
	\end{center}
\end{table*}

\begin{figure}[H]
	\begin{minipage}[t]{0.5\linewidth}
		\centering
        \includegraphics[width=0.90\linewidth, height=3cm]{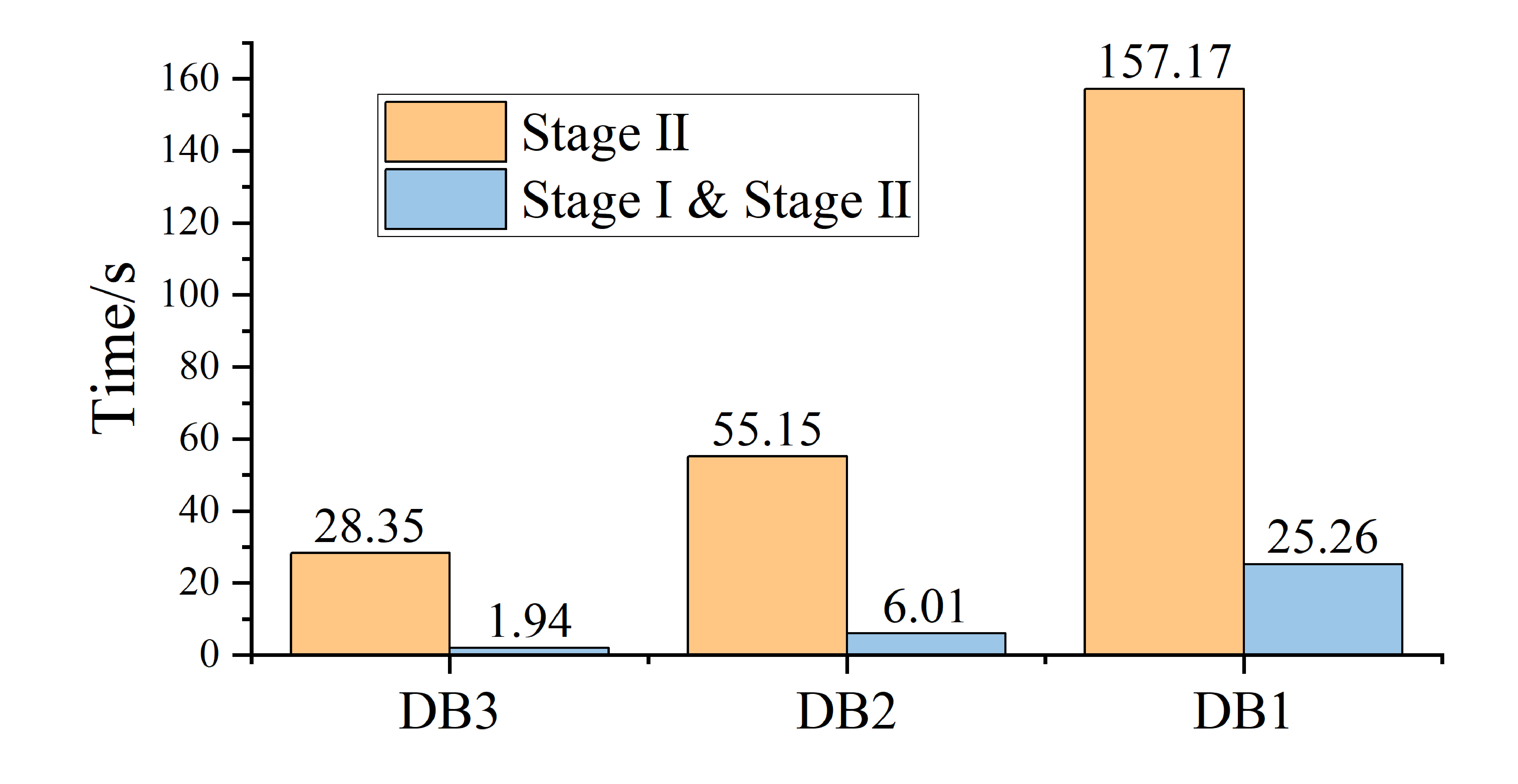}
        \caption{Impact of pruning}
        \label{fig:RQ3_time}
	\end{minipage}%
	\begin{minipage}[t]{0.5\linewidth}
		\centering
        \includegraphics[width=0.90\linewidth, height=3cm]{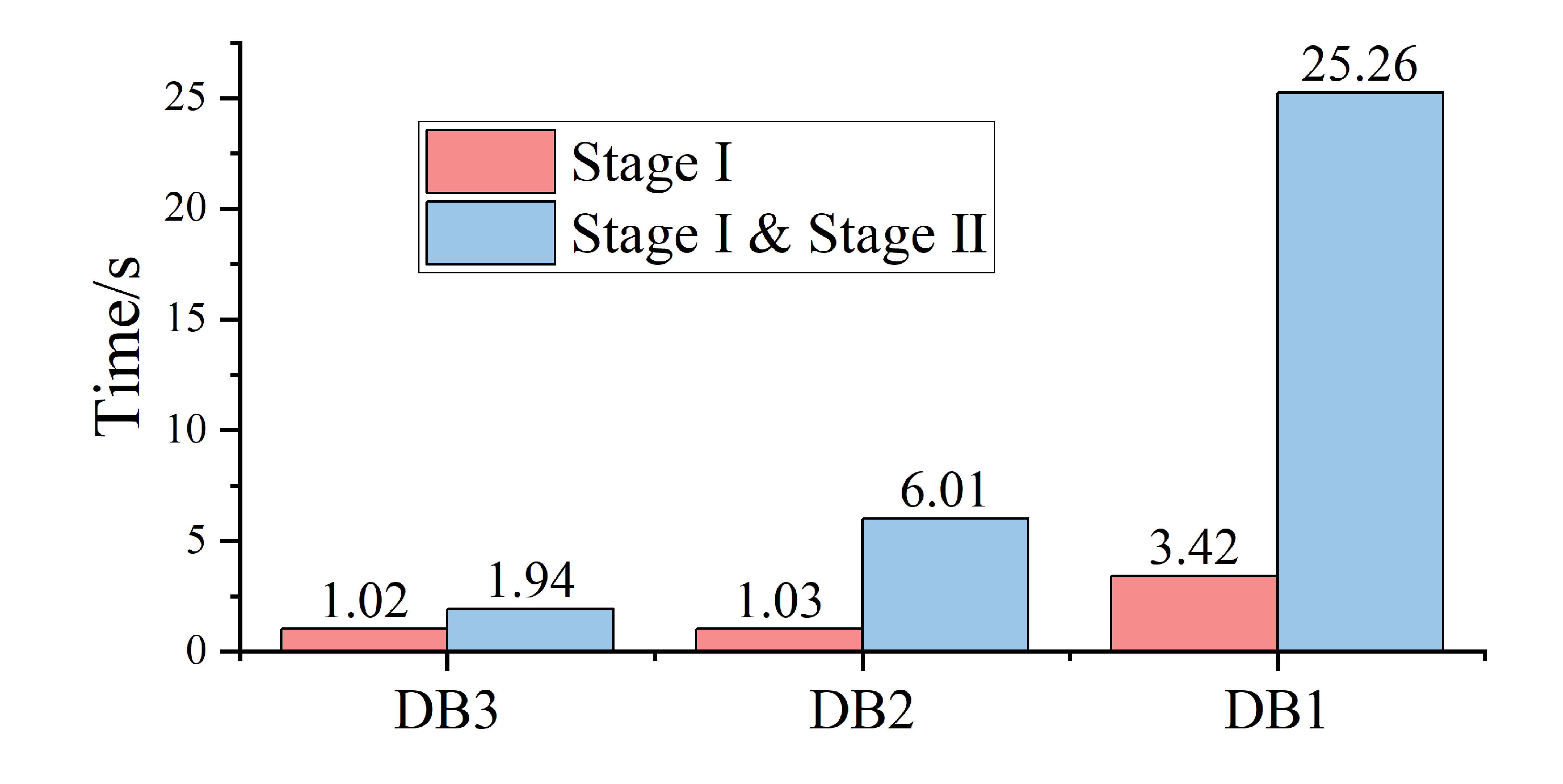}
        \caption{Impact of symbolic execution verification}
        \label{fig:RQ4_time}
	\end{minipage}
\end{figure}

We also evaluated the impact of pruning in Stage \uppercase\expandafter{\romannumeral1} on the overall detection time, as shown in Figure~\ref{fig:RQ3_time}. Pruning significantly reduces the execution time. For complex contract vulnerability detection (DB1), the average detection time decreases from 157.17s to 25.26s, greatly improving the efficiency of detecting vulnerabilities in complex contracts. This reduction in execution time highlights the importance of precise pruning in Stage \uppercase\expandafter{\romannumeral1} for efficient vulnerability detection.

\vspace{-0.1cm}
\vspace{-0.2cm}


\textbf{Answer to RQ3:} Pruning in Stage \uppercase\expandafter{\romannumeral1} significantly improves detection accuracy, boosting the F1 score from 6.59\% to 78.65\%. Additionally, the precise path pruning in Stage \uppercase\expandafter{\romannumeral1} ensures the efficiency of vulnerability detection.

\subsection{Impact of Symbolic Execution Verification in Stage \uppercase\expandafter{\romannumeral2}}

In Stage \uppercase\expandafter{\romannumeral2}, SliSE uses symbolic execution to validate the reachability of suspicious vulnerability paths, reducing false positives in the detection results. To evaluate the impact of symbolic execution verification in Stage \uppercase\expandafter{\romannumeral2} for overall detection, we conducted comparative experiments by comparing the results between \textit{Stage \uppercase\expandafter{\romannumeral1}} and \textit{Stage \uppercase\expandafter{\romannumeral1} \& \uppercase\expandafter{\romannumeral2}} modes (representing without and with Stage \uppercase\expandafter{\romannumeral2}). Table~\ref{tab:RQ2} presents the statistics for Stage \uppercase\expandafter{\romannumeral1} and Stage \uppercase\expandafter{\romannumeral1} \& \uppercase\expandafter{\romannumeral2}. The data clearly show that Stage \uppercase\expandafter{\romannumeral2} symbolic execution verification significantly reduces the number of false positives (FPs) in the results. For Reentrancy vulnerability detection on DB1 only with Stage \uppercase\expandafter{\romannumeral1}, there were 78 FPs. When combined with Stage \uppercase\expandafter{\romannumeral2} symbolic execution verification, the number of FPs notably decreased, resulting in a substantial increase in precision from 47.30\% to 72.16\%. This highlights that the symbolic execution verification in Stage \uppercase\expandafter{\romannumeral2} ensures more precise vulnerability detection results.

We also analyzed the time overhead associated with Stage \uppercase\expandafter{\romannumeral2}'s symbolic execution verification. As shown in Figure~\ref{fig:RQ4_time}, the overhead introduced by symbolic execution verification is less than 7 seconds on DB2, DB3, with the average detection time remaining at 25.26 seconds for DB1. This time investment is significantly lower compared to the runtime of most symbolic execution tools and is suitable for batch detection purposes. 

\textbf{Answer to RQ4:} Symbolic execution verification in Stage \uppercase\expandafter{\romannumeral2} significantly reduces FPs, boosting precision from 47.30\% to 72.16\% on DB1. Moreover, symbolic execution verification incurs some time overhead, the overall detection time remains reasonable, ensuring efficient detection processes.

\vspace{-0.2cm}
\section{Discussion}
\label{sec:discussion}

\subsection{Capability Analysis of Existing Tools}

Zheng et al.~\cite{zheng2023dappscan} conducted an analysis of 1,322 open-source DApps audit reports from 30 security audit teams, compiling a large-scale, fair, real-world smart contract vulnerability dataset known as DB1. Their experiments revealed that existing tools performed poorly in terms of effectiveness and successful detection rates, suggesting that future development should prioritize real-world complex contracts over simple toy contracts. To assess the capability of existing tools to detect complex contracts, we quantified the complexity of these contracts and conducted comparative analysis.

\vspace{-0.15cm}
\begin{table}[htb]
    \begin{center}
        \caption{Detection capability of existing tools for complex contracts}
        \label{tab:dicussion}
        \begin{threeparttable}
            \resizebox{0.88\columnwidth}{!}{
            \large 
\begin{tabular}{l|ll|ll|ll|ll|ll}
\hline
       & \multicolumn{2}{l|}{AvgCyclomatic} & \multicolumn{2}{l|}{MaxCyclomatic} & \multicolumn{2}{l|}{SumCyclomatic} & \multicolumn{2}{l|}{MaxNesting} & \multicolumn{2}{l}{CountContractCoupled} \\ \cline{2-11} 
       & Detected      & Avg.     & Detected      & Avg.     & Detected      & Avg.     & Detected    & Avg.    & Detected         & Avg.        \\ \hline
Mean   & 1.3(36.1\%)          & 3.6         & 4.0(18.7\%)          & 21.4        & 39.5(15.6\%)         & 253.5       & 1.8(75.0\%)        & 2.4        & 0.8(34.8\%)             & 2.3            \\
Max    & 2.0(25.3\%)          & 7.9         & 11.0(14.9\%)         & 74.0        & 104.0(11.0\%)        & 949.0       & 4.0(33.3\%)        & 12.0       & 3.0(25.0\%)             & 12.0           \\
Median & 1.2(36.4\%)          & 3.3         & 2.0(12.1\%)          & 16.5        & 25.0(14.6\%)         & 171.0       & 1.0(50.0\%)        & 2.0        & 0.0(0.0\%)              & 2.0            \\
Std.   & 0.3(20.0\%)          & 1.5         & 3.2(19.8\%)          & 16.2        & 34.2(14.5\%)         & 236.1       & 1.5(88.2\%)        & 1.7        & 1.3(48.1\%)             & 2.7            \\ \hline
\end{tabular}
            }
            \begin{tablenotes}
            \tiny
            \item[] 
            \hspace{0.5cm} * \textit{Detected} means successfully detected by existing tool, \textit{Avg.} means the overall average complexity.
		    \end{tablenotes}
        \end{threeparttable}
    \end{center}
\end{table}
\vspace{-0.20cm}

We utilized the \textit{Complexity Metrics} proposed by Chao et al.~\cite{complexity} to quantify the complexity of contracts and assess the detection capabilities of existing tools on complex contracts. These metrics provide insights into the complexity and interdependencies of functions and contracts. Specifically, \textit{AvgCyclomatic} calculates the average cyclomatic complexity among functions, \textit{MaxCyclomatic} identifies the most complex function, and \textit{SumCyclomatic} aggregates complexities across functions. \textit{MaxNesting} reveals the deepest nesting of control structures in functions, while \textit{CountContractCoupled} counts interconnected contracts, indicating their dependencies. We compared the average complexity metrics of contracts successfully detected by existing tools with the overall average complexity metrics of DB1, as shown in Table~\ref{tab:dicussion}. The data in the table reveals that only around 30\% of the overall contract complexity can be successfully detected by these tools. This highlights the limited capabilities of existing tools in identifying vulnerabilities in real-world complex contracts. They are insufficient for addressing the security needs of practical contracts in DApps, indicating a significant challenge in detecting Reentrancy vulnerabilities in complex contracts.

\subsection{Threats to Validity}
\noindent{\bf External Validity.} 
We validated SliSE's Reentrancy vulnerability detection performance on Ethereum using the DB2 and DB3 datasets. DB2 collected from positive labels in existing tool detection results on Ethereum contracts, without considering negative labels. However, labeling all Ethereum contracts is time-consuming and error-prone, which demands a substantial team of experienced engineers for Reentrancy vulnerability analysis. DB2 was created by subjecting 230,548 verified smart contracts from Etherscan to scans by five automated detection tools~\cite{9mythril,luu2016making,tsankov2018securify,choi2021smartian,bose2022sailfish}, followed by two rounds of manual examination, Zheng et al. underscored the error-prone nature of vulnerability labeling~\cite{turntherudder}. In addition, DB3 is the well-known SmartBugs dataset~\cite{8Smartbugs}. These datasets are widely acknowledged and have been openly shared at top software engineering conferences, establishing them as reliable sources of ground truth for Ethereum contracts.

\noindent{\bf Internal Validity.} 
While various permission controls in smart contracts are essential for security, they present a challenge that demands specific semantic comprehension to assess path feasibility. SliSE does not account for these specific semantic requirements, resulting in false positives in detection results. To expand the scope of vulnerability detection, we draw inspiration from a range of patterns in \textit{path protective techniques} (PPTs)~\cite{11ye2020clairvoyance}. These patterns have been identified through empirical research and help identify common false positives in state-of-the-art tool rules. By applying these patterns, we effectively reduce false positives, thereby improving scalability and effectiveness.

\vspace{-0.25cm}
\section{Related Work}
\label{sec:relatedwork}

\subsection{Reentrancy Vulnerability Detection}
There are numerous tools that support Reentrancy vulnerability detection, utilizing various vulnerability detection techniques, including static analysis, symbolic execution, fuzz testing, and formal verification. Tools like Oyente~\cite{luu2016making}, Osiris~\cite{13torres2018osiris} utilize  CFG-based bytecode analysis with pattern matching for fast and scalable Reentrancy vulnerability detection. Slither~\cite{feist2019slither} generates intermediate language \textit{SlithIR} by analyzing smart contract AST, employing rule-based detection for vulnerabilities, and scaling up to 5 types of Reentrancy vulnerabilities. Clairvoyance~\cite{11ye2020clairvoyance} introduces taint analysis via shadow stacks to reduce false positives while combining lightweight symbolic execution for vulnerability detection. Mythril~\cite{9mythril}, Manticore~\cite{29mossberg2019manticore} use emulation testing to generate sequences of transactions triggering vulnerabilities. MPro~\cite{24zhang2019mpro} simulates execution by increasing the depth of symbolic execution analysis, exploring deeper levels of the state space. Smartian~\cite{choi2021smartian} applies data flow analysis to guide path exploration, generating critical transaction sequences triggering vulnerabilities. Sailfish~\cite{bose2022sailfish} analyzes program dependency information to generate \textit{Storage Dependency Graph} (SDG) detects dangerous access patterns, but can not analyze global cross-contract data flow. Pluto~\cite{14ma2021pluto} supports cross-contract vulnerability detection by constructing \textit{Inter-Contract Control Flow Graph} (I-CFG) to enable cross-contract bytecode traversal, then exploring I-CFG to detect vulnerabilities. Park~\cite{zheng2022park} proposed a method based on parallel symbolic execution, which proposed a dynamic forking algorithm based on process forking to speed up vulnerability detection.

\textbf{Differing from existing methods.} 
SliSE constructs the \textit{Inter-contract Program Dependency Graph} (I-PDG) to provide global control and data dependencies within contracts, including cross-contract data flows. In contrast, Sailfish~\cite{bose2022sailfish} exclusively conducts program dependency analysis on state variables within single contract and does not support cross-contract analysis. EtherSolve~\cite{contro2021ethersolve}  calculates dynamic jump address using the \textit{symbolic stack}, which always fails in complex contracts due to issues with excessive recursion depth. SliSE leverages constant propagation analysis with SSA to quickly compute dynamic jump addresses and recovery the CFG paths. Compared to Pluto~\cite{14ma2021pluto}, SliSE primarily slices out code blocks related to Reentrancy vulnerabilities from complex contracts for focused analysis, employing symbolic execution for efficient and reliable verification. Evaluation results show that SliSE outperforms many state-of-the-art tools in detecting Reentrancy vulnerabilities in complex contracts.

\vspace{-0.25cm}
\section{Conclusion}
\label{sec:conclusion}
In this paper, we introduce a tool named SliSE, designed for efficient detection of Reentrancy vulnerabilities in complex contracts. The detection process consists of two stages: Warning Search and Symbolic Execution Verification. In the Warning Search stage, SliSE analyzes the \textit{Inter-contract Program Dependency Graph} (I-PDG) with program slicing, collecting suspicious vulnerability information as warnings. In the Symbolic Execution Verification stage, it employs symbolic execution to traverse the paths indicated by the warning information and validate their reachability, ensuring effective vulnerability detection. In comparative experiments, SliSE achieves impressive results, with an F1 score of 78.65\%, surpassing the highest score of 9.26\% achieved by eight existing state-of-the-art tools. Additionally, it achieves a recall rate exceeding 90\% for Reentrancy vulnerability detection of  contracts on Ethereum. Overall, SliSE provides an effective solution for detecting reentrancy vulnerabilities in complex contracts.


\begin{acks}
The work described in this paper is supported by the National Natural Science Foundation of China (62032025, 62302534, 62332004), and the Major Key Project of Peng Cheng Laboratory under Grant PCL2023A05-2.
\end{acks}


\bibliographystyle{ACM-Reference-Format}
\bibliography{ref}


\begin{thebibliography}{48}


\ifx \showCODEN    \undefined \def \showCODEN     #1{\unskip}     \fi
\ifx \showDOI      \undefined \def \showDOI       #1{#1}\fi
\ifx \showISBNx    \undefined \def \showISBNx     #1{\unskip}     \fi
\ifx \showISBNxiii \undefined \def \showISBNxiii  #1{\unskip}     \fi
\ifx \showISSN     \undefined \def \showISSN      #1{\unskip}     \fi
\ifx \showLCCN     \undefined \def \showLCCN      #1{\unskip}     \fi
\ifx \shownote     \undefined \def \shownote      #1{#1}          \fi
\ifx \showarticletitle \undefined \def \showarticletitle #1{#1}   \fi
\ifx \showURL      \undefined \def \showURL       {\relax}        \fi
\providecommand\bibfield[2]{#2}
\providecommand\bibinfo[2]{#2}
\providecommand\natexlab[1]{#1}
\providecommand\showeprint[2][]{arXiv:#2}

\bibitem[\protect\citeauthoryear{Bose, Das, Chen, Feng, Kruegel, and
  Vigna}{Bose et~al\mbox{.}}{2022}]%
        {bose2022sailfish}
\bibfield{author}{\bibinfo{person}{Priyanka Bose}, \bibinfo{person}{Dipanjan
  Das}, \bibinfo{person}{Yanju Chen}, \bibinfo{person}{Yu Feng},
  \bibinfo{person}{Christopher Kruegel}, {and} \bibinfo{person}{Giovanni
  Vigna}.} \bibinfo{year}{2022}\natexlab{}.
\newblock \showarticletitle{SAILFISH: Vetting Smart Contract
  State-Inconsistency Bugs in Seconds}. In \bibinfo{booktitle}{\emph{2022 IEEE
  Symposium on Security and Privacy (SP)}}. \bibinfo{pages}{161--178}.
\newblock
\urldef\tempurl%
\url{https://doi.org/10.1109/SP46214.2022.9833721}
\showDOI{\tempurl}


\bibitem[\protect\citeauthoryear{Canfora, Cimitile, and De~Lucia}{Canfora
  et~al\mbox{.}}{1998}]%
        {canfora1998conditioned}
\bibfield{author}{\bibinfo{person}{Gerardo Canfora}, \bibinfo{person}{Aniello
  Cimitile}, {and} \bibinfo{person}{Andrea De~Lucia}.}
  \bibinfo{year}{1998}\natexlab{}.
\newblock \showarticletitle{Conditioned program slicing}.
\newblock \bibinfo{journal}{\emph{Information and Software Technology}}
  \bibinfo{volume}{40}, \bibinfo{number}{11-12} (\bibinfo{year}{1998}),
  \bibinfo{pages}{595--607}.
\newblock


\bibitem[\protect\citeauthoryear{Cecchetti, Yao, Ni, and Myers}{Cecchetti
  et~al\mbox{.}}{2021}]%
        {compositional_reentrancy}
\bibfield{author}{\bibinfo{person}{Ethan Cecchetti}, \bibinfo{person}{Siqiu
  Yao}, \bibinfo{person}{Haobin Ni}, {and} \bibinfo{person}{Andrew~C. Myers}.}
  \bibinfo{year}{2021}\natexlab{}.
\newblock \showarticletitle{Compositional Security for Reentrant Applications}.
  In \bibinfo{booktitle}{\emph{2021 IEEE Symposium on Security and Privacy
  (SP)}}. \bibinfo{pages}{1249--1267}.
\newblock
\urldef\tempurl%
\url{https://doi.org/10.1109/SP40001.2021.00084}
\showDOI{\tempurl}


\bibitem[\protect\citeauthoryear{Chaliasos, Charalambous, Zhou, Galanopoulou,
  Gervais, Mitropoulos, and Livshits}{Chaliasos et~al\mbox{.}}{2023}]%
        {chaliasos2023smart}
\bibfield{author}{\bibinfo{person}{Stefanos Chaliasos},
  \bibinfo{person}{Marcos~Antonios Charalambous}, \bibinfo{person}{Liyi Zhou},
  \bibinfo{person}{Rafaila Galanopoulou}, \bibinfo{person}{Arthur Gervais},
  \bibinfo{person}{Dimitris Mitropoulos}, {and} \bibinfo{person}{Ben
  Livshits}.} \bibinfo{year}{2023}\natexlab{}.
\newblock \showarticletitle{Smart contract and defi security: Insights from
  tool evaluations and practitioner surveys}.
\newblock \bibinfo{journal}{\emph{arXiv preprint arXiv:2304.02981}}
  (\bibinfo{year}{2023}).
\newblock


\bibitem[\protect\citeauthoryear{Chen, Huang, Lin, Zheng, and Zheng}{Chen
  et~al\mbox{.}}{2023}]%
        {swe}
\bibfield{author}{\bibinfo{person}{Jiachi Chen}, \bibinfo{person}{Mingyuan
  Huang}, \bibinfo{person}{Zewei Lin}, \bibinfo{person}{Peilin Zheng}, {and}
  \bibinfo{person}{Zibin Zheng}.} \bibinfo{year}{2023}\natexlab{}.
\newblock \bibinfo{title}{To Healthier Ethereum: A Comprehensive and Iterative
  Smart Contract Weakness Enumeration}.
\newblock
\newblock
\showeprint[arxiv]{cs.SE/2308.10227}


\bibitem[\protect\citeauthoryear{Chen, Xia, Lo, Grundy, Luo, and Chen}{Chen
  et~al\mbox{.}}{2022}]%
        {chen2022defining}
\bibfield{author}{\bibinfo{person}{Jiachi Chen}, \bibinfo{person}{Xin Xia},
  \bibinfo{person}{David Lo}, \bibinfo{person}{John Grundy},
  \bibinfo{person}{Xiapu Luo}, {and} \bibinfo{person}{Ting Chen}.}
  \bibinfo{year}{2022}\natexlab{}.
\newblock \showarticletitle{Defining Smart Contract Defects on Ethereum}.
\newblock \bibinfo{journal}{\emph{IEEE Transactions on Software Engineering}}
  \bibinfo{volume}{48}, \bibinfo{number}{1} (\bibinfo{year}{2022}),
  \bibinfo{pages}{327--345}.
\newblock
\urldef\tempurl%
\url{https://doi.org/10.1109/TSE.2020.2989002}
\showDOI{\tempurl}


\bibitem[\protect\citeauthoryear{Chen, Zhang, Li, Luo, Wang, Cao, Xiao, and
  Zhang}{Chen et~al\mbox{.}}{2019}]%
        {chen2019tokenscope}
\bibfield{author}{\bibinfo{person}{Ting Chen}, \bibinfo{person}{Yufei Zhang},
  \bibinfo{person}{Zihao Li}, \bibinfo{person}{Xiapu Luo},
  \bibinfo{person}{Ting Wang}, \bibinfo{person}{Rong Cao},
  \bibinfo{person}{Xiuzhuo Xiao}, {and} \bibinfo{person}{Xiaosong Zhang}.}
  \bibinfo{year}{2019}\natexlab{}.
\newblock \showarticletitle{TokenScope: Automatically Detecting Inconsistent
  Behaviors of Cryptocurrency Tokens in Ethereum}. In
  \bibinfo{booktitle}{\emph{Proceedings of the 2019 ACM SIGSAC Conference on
  Computer and Communications Security}} \emph{(\bibinfo{series}{CCS '19})}.
  \bibinfo{publisher}{Association for Computing Machinery},
  \bibinfo{address}{New York, NY, USA}, \bibinfo{pages}{1503–1520}.
\newblock
\showISBNx{9781450367479}
\urldef\tempurl%
\url{https://doi.org/10.1145/3319535.3345664}
\showDOI{\tempurl}


\bibitem[\protect\citeauthoryear{Choi, Kim, Kim, Grieco, Groce, and Cha}{Choi
  et~al\mbox{.}}{2021}]%
        {choi2021smartian}
\bibfield{author}{\bibinfo{person}{Jaeseung Choi}, \bibinfo{person}{Doyeon
  Kim}, \bibinfo{person}{Soomin Kim}, \bibinfo{person}{Gustavo Grieco},
  \bibinfo{person}{Alex Groce}, {and} \bibinfo{person}{Sang~Kil Cha}.}
  \bibinfo{year}{2021}\natexlab{}.
\newblock \showarticletitle{SMARTIAN: Enhancing Smart Contract Fuzzing with
  Static and Dynamic Data-Flow Analyses}. In \bibinfo{booktitle}{\emph{2021
  36th IEEE/ACM International Conference on Automated Software Engineering
  (ASE)}}. \bibinfo{pages}{227--239}.
\newblock
\urldef\tempurl%
\url{https://doi.org/10.1109/ASE51524.2021.9678888}
\showDOI{\tempurl}


\bibitem[\protect\citeauthoryear{ConsenSys}{ConsenSys}{2020}]%
        {9mythril}
\bibfield{author}{\bibinfo{person}{ConsenSys}.}
  \bibinfo{year}{2020}\natexlab{}.
\newblock \bibinfo{title}{Mythril}.
\newblock
\newblock
\urldef\tempurl%
\url{https://github.com/ConsenSys/mythril}
\showURL{%
\tempurl}


\bibitem[\protect\citeauthoryear{Contro, Crosara, Ceccato, and
  Dalla~Preda}{Contro et~al\mbox{.}}{2021}]%
        {contro2021ethersolve}
\bibfield{author}{\bibinfo{person}{Filippo Contro}, \bibinfo{person}{Marco
  Crosara}, \bibinfo{person}{Mariano Ceccato}, {and} \bibinfo{person}{Mila
  Dalla~Preda}.} \bibinfo{year}{2021}\natexlab{}.
\newblock \showarticletitle{Ethersolve: Computing an accurate control-flow
  graph from ethereum bytecode}. In \bibinfo{booktitle}{\emph{2021 IEEE/ACM
  29th International Conference on Program Comprehension (ICPC)}}. IEEE,
  \bibinfo{pages}{127--137}.
\newblock


\bibitem[\protect\citeauthoryear{Daian}{Daian}{2016}]%
        {DAO}
\bibfield{author}{\bibinfo{person}{Phil Daian}.}
  \bibinfo{year}{2016}\natexlab{}.
\newblock \bibinfo{title}{Analysis of the DAO exploit}.
\newblock
\newblock
\urldef\tempurl%
\url{https://hackingdistributed.com/2016/06/18/analysis-of-the-dao-exploit/}
\showURL{%
\tempurl}


\bibitem[\protect\citeauthoryear{Feist, Grieco, and Groce}{Feist
  et~al\mbox{.}}{2019}]%
        {feist2019slither}
\bibfield{author}{\bibinfo{person}{Josselin Feist}, \bibinfo{person}{Gustavo
  Grieco}, {and} \bibinfo{person}{Alex Groce}.}
  \bibinfo{year}{2019}\natexlab{}.
\newblock \showarticletitle{Slither: a static analysis framework for smart
  contracts}. In \bibinfo{booktitle}{\emph{2019 IEEE/ACM 2nd International
  Workshop on Emerging Trends in Software Engineering for Blockchain
  (WETSEB)}}. IEEE, \bibinfo{pages}{8--15}.
\newblock


\bibitem[\protect\citeauthoryear{Ghaleb, Rubin, and Pattabiraman}{Ghaleb
  et~al\mbox{.}}{2022}]%
        {ghaleb2022etainter}
\bibfield{author}{\bibinfo{person}{Asem Ghaleb}, \bibinfo{person}{Julia Rubin},
  {and} \bibinfo{person}{Karthik Pattabiraman}.}
  \bibinfo{year}{2022}\natexlab{}.
\newblock \showarticletitle{eTainter: detecting gas-related vulnerabilities in
  smart contracts}. In \bibinfo{booktitle}{\emph{Proceedings of the 31st ACM
  SIGSOFT International Symposium on Software Testing and Analysis}}.
  \bibinfo{pages}{728--739}.
\newblock


\bibitem[\protect\citeauthoryear{Go}{Go}{2018}]%
        {CEI2}
\bibfield{author}{\bibinfo{person}{Seungwon Go}.}
  \bibinfo{year}{2018}\natexlab{}.
\newblock \bibinfo{title}{Smart Contract : Security Patterns}.
\newblock
\newblock
\urldef\tempurl%
\url{https://medium.com/returnvalues/smart-contract-security-patterns-79e03b5a1659}
\showURL{%
\tempurl}


\bibitem[\protect\citeauthoryear{Han, Li, Zhang, Xu, et~al\mbox{.}}{Han
  et~al\mbox{.}}{2023}]%
        {han2023smart}
\bibfield{author}{\bibinfo{person}{Daojun Han}, \bibinfo{person}{Qiuyue Li},
  \bibinfo{person}{Lei Zhang}, \bibinfo{person}{Tao Xu}, {et~al\mbox{.}}}
  \bibinfo{year}{2023}\natexlab{}.
\newblock \showarticletitle{A Smart Contract Vulnerability Detection Model
  Based on Syntactic and Semantic Fusion Learning}.
\newblock \bibinfo{journal}{\emph{Wireless Communications and Mobile
  Computing}}  \bibinfo{volume}{2023} (\bibinfo{year}{2023}).
\newblock


\bibitem[\protect\citeauthoryear{Harrold, Malloy, and Rothermel}{Harrold
  et~al\mbox{.}}{1993}]%
        {harrold1993efficient}
\bibfield{author}{\bibinfo{person}{Mary~Jean Harrold}, \bibinfo{person}{Brian
  Malloy}, {and} \bibinfo{person}{Gregg Rothermel}.}
  \bibinfo{year}{1993}\natexlab{}.
\newblock \showarticletitle{Efficient construction of program dependence
  graphs}.
\newblock \bibinfo{journal}{\emph{ACM SIGSOFT Software Engineering Notes}}
  \bibinfo{volume}{18}, \bibinfo{number}{3} (\bibinfo{year}{1993}),
  \bibinfo{pages}{160--170}.
\newblock


\bibitem[\protect\citeauthoryear{He, Balunovi\'{c}, Ambroladze, Tsankov, and
  Vechev}{He et~al\mbox{.}}{2019}]%
        {he2019learning}
\bibfield{author}{\bibinfo{person}{Jingxuan He}, \bibinfo{person}{Mislav
  Balunovi\'{c}}, \bibinfo{person}{Nodar Ambroladze}, \bibinfo{person}{Petar
  Tsankov}, {and} \bibinfo{person}{Martin Vechev}.}
  \bibinfo{year}{2019}\natexlab{}.
\newblock \showarticletitle{Learning to Fuzz from Symbolic Execution with
  Application to Smart Contracts}. In \bibinfo{booktitle}{\emph{Proceedings of
  the 2019 ACM SIGSAC Conference on Computer and Communications Security}}
  \emph{(\bibinfo{series}{CCS '19})}. \bibinfo{publisher}{Association for
  Computing Machinery}, \bibinfo{address}{New York, NY, USA},
  \bibinfo{pages}{531–548}.
\newblock
\showISBNx{9781450367479}
\urldef\tempurl%
\url{https://doi.org/10.1145/3319535.3363230}
\showDOI{\tempurl}


\bibitem[\protect\citeauthoryear{Holler, Biewer, and Schneidewind}{Holler
  et~al\mbox{.}}{2023}]%
        {holler2023horstify}
\bibfield{author}{\bibinfo{person}{Sebastian Holler},
  \bibinfo{person}{Sebastian Biewer}, {and} \bibinfo{person}{Clara
  Schneidewind}.} \bibinfo{year}{2023}\natexlab{}.
\newblock \showarticletitle{HoRStify: Sound Security Analysis of Smart
  Contracts}.
\newblock \bibinfo{journal}{\emph{arXiv preprint arXiv:2301.13769}}
  (\bibinfo{year}{2023}).
\newblock


\bibitem[\protect\citeauthoryear{insurgent}{insurgent}{2022}]%
        {CEI1}
\bibfield{author}{\bibinfo{person}{insurgent}.}
  \bibinfo{year}{2022}\natexlab{}.
\newblock \bibinfo{title}{Solidity Smart Contract Security: 4 Ways to Prevent
  Reentrancy Attacks}.
\newblock
\newblock
\urldef\tempurl%
\url{https://betterprogramming.pub/solidity-smart-contract-security-preventing-reentrancy-attacks-fc729339a3ff}
\showURL{%
\tempurl}


\bibitem[\protect\citeauthoryear{Jacques~Dafflon}{Jacques~Dafflon}{2017}]%
        {ERC777}
\bibfield{author}{\bibinfo{person}{Thomas~Shababi Jacques~Dafflon,
  Jordi~Baylina}.} \bibinfo{year}{2017}\natexlab{}.
\newblock \bibinfo{title}{ERC-777: Token Standard}.
\newblock
\newblock
\urldef\tempurl%
\url{https://eips.ethereum.org/EIPS/eip-777}
\showURL{%
\tempurl}


\bibitem[\protect\citeauthoryear{Jiang, Liu, and Chan}{Jiang
  et~al\mbox{.}}{2018}]%
        {jiang2018contractfuzzer}
\bibfield{author}{\bibinfo{person}{Bo Jiang}, \bibinfo{person}{Ye Liu}, {and}
  \bibinfo{person}{W.~K. Chan}.} \bibinfo{year}{2018}\natexlab{}.
\newblock \showarticletitle{ContractFuzzer: Fuzzing Smart Contracts for
  Vulnerability Detection}. In \bibinfo{booktitle}{\emph{Proceedings of the
  33rd ACM/IEEE International Conference on Automated Software Engineering}}
  \emph{(\bibinfo{series}{ASE '18})}. \bibinfo{publisher}{Association for
  Computing Machinery}, \bibinfo{address}{New York, NY, USA},
  \bibinfo{pages}{259–269}.
\newblock
\showISBNx{9781450359375}
\urldef\tempurl%
\url{https://doi.org/10.1145/3238147.3238177}
\showDOI{\tempurl}


\bibitem[\protect\citeauthoryear{Kalra, Goel, Dhawan, and Sharma}{Kalra
  et~al\mbox{.}}{2018}]%
        {kalra2018zeus}
\bibfield{author}{\bibinfo{person}{Sukrit Kalra}, \bibinfo{person}{Seep Goel},
  \bibinfo{person}{Mohan Dhawan}, {and} \bibinfo{person}{Subodh Sharma}.}
  \bibinfo{year}{2018}\natexlab{}.
\newblock \showarticletitle{Zeus: analyzing safety of smart contracts.}. In
  \bibinfo{booktitle}{\emph{Ndss}}. \bibinfo{pages}{1--12}.
\newblock


\bibitem[\protect\citeauthoryear{Krupp and Rossow}{Krupp and Rossow}{2018}]%
        {krupp2018teether}
\bibfield{author}{\bibinfo{person}{Johannes Krupp} {and}
  \bibinfo{person}{Christian Rossow}.} \bibinfo{year}{2018}\natexlab{}.
\newblock \showarticletitle{{teEther: Gnawing at Ethereum to Automatically
  Exploit Smart Contracts}}. In \bibinfo{booktitle}{\emph{27th {USENIX}
  Security Symposium ({USENIX} Security 18)}}. \bibinfo{pages}{1317--1333}.
\newblock


\bibitem[\protect\citeauthoryear{Liao, Zheng, Chen, and Nan}{Liao
  et~al\mbox{.}}{2022}]%
        {liao2022smartdagger}
\bibfield{author}{\bibinfo{person}{Zeqin Liao}, \bibinfo{person}{Zibin Zheng},
  \bibinfo{person}{Xiao Chen}, {and} \bibinfo{person}{Yuhong Nan}.}
  \bibinfo{year}{2022}\natexlab{}.
\newblock \showarticletitle{SmartDagger: A Bytecode-Based Static Analysis
  Approach for Detecting Cross-Contract Vulnerability}. In
  \bibinfo{booktitle}{\emph{Proceedings of the 31st ACM SIGSOFT International
  Symposium on Software Testing and Analysis}} \emph{(\bibinfo{series}{ISSTA
  2022})}. \bibinfo{publisher}{Association for Computing Machinery},
  \bibinfo{address}{New York, NY, USA}, \bibinfo{pages}{752–764}.
\newblock
\showISBNx{9781450393799}
\urldef\tempurl%
\url{https://doi.org/10.1145/3533767.3534222}
\showDOI{\tempurl}


\bibitem[\protect\citeauthoryear{Luu, Chu, Olickel, Saxena, and Hobor}{Luu
  et~al\mbox{.}}{2016}]%
        {luu2016making}
\bibfield{author}{\bibinfo{person}{Loi Luu}, \bibinfo{person}{Duc-Hiep Chu},
  \bibinfo{person}{Hrishi Olickel}, \bibinfo{person}{Prateek Saxena}, {and}
  \bibinfo{person}{Aquinas Hobor}.} \bibinfo{year}{2016}\natexlab{}.
\newblock \showarticletitle{Making Smart Contracts Smarter}. In
  \bibinfo{booktitle}{\emph{Proceedings of the 2016 ACM SIGSAC Conference on
  Computer and Communications Security}} \emph{(\bibinfo{series}{CCS '16})}.
  \bibinfo{publisher}{Association for Computing Machinery},
  \bibinfo{address}{New York, NY, USA}, \bibinfo{pages}{254–269}.
\newblock
\showISBNx{9781450341394}
\urldef\tempurl%
\url{https://doi.org/10.1145/2976749.2978309}
\showDOI{\tempurl}


\bibitem[\protect\citeauthoryear{Ma, Xu, Ren, Yin, Chen, Qiao, Gu, Li, Jiang,
  and Sun}{Ma et~al\mbox{.}}{2021}]%
        {14ma2021pluto}
\bibfield{author}{\bibinfo{person}{Fuchen Ma}, \bibinfo{person}{Zhenyang Xu},
  \bibinfo{person}{Meng Ren}, \bibinfo{person}{Zijing Yin},
  \bibinfo{person}{Yuanliang Chen}, \bibinfo{person}{Lei Qiao},
  \bibinfo{person}{Bin Gu}, \bibinfo{person}{Huizhong Li}, \bibinfo{person}{Yu
  Jiang}, {and} \bibinfo{person}{Jiaguang Sun}.}
  \bibinfo{year}{2021}\natexlab{}.
\newblock \showarticletitle{Pluto: Exposing vulnerabilities in inter-contract
  scenarios}.
\newblock \bibinfo{journal}{\emph{IEEE Transactions on Software Engineering}}
  \bibinfo{volume}{48}, \bibinfo{number}{11} (\bibinfo{year}{2021}),
  \bibinfo{pages}{4380--4396}.
\newblock


\bibitem[\protect\citeauthoryear{Mossberg, Manzano, Hennenfent, Groce, Grieco,
  Feist, Brunson, and Dinaburg}{Mossberg et~al\mbox{.}}{2019}]%
        {29mossberg2019manticore}
\bibfield{author}{\bibinfo{person}{Mark Mossberg}, \bibinfo{person}{Felipe
  Manzano}, \bibinfo{person}{Eric Hennenfent}, \bibinfo{person}{Alex Groce},
  \bibinfo{person}{Gustavo Grieco}, \bibinfo{person}{Josselin Feist},
  \bibinfo{person}{Trent Brunson}, {and} \bibinfo{person}{Artem Dinaburg}.}
  \bibinfo{year}{2019}\natexlab{}.
\newblock \showarticletitle{Manticore: A user-friendly symbolic execution
  framework for binaries and smart contracts}. In
  \bibinfo{booktitle}{\emph{2019 34th IEEE/ACM International Conference on
  Automated Software Engineering (ASE)}}. IEEE, \bibinfo{pages}{1186--1189}.
\newblock


\bibitem[\protect\citeauthoryear{Nguyen, Pham, Sun, Lin, and Minh}{Nguyen
  et~al\mbox{.}}{2020}]%
        {nguyen2020sfuzz}
\bibfield{author}{\bibinfo{person}{Tai~D. Nguyen}, \bibinfo{person}{Long~H.
  Pham}, \bibinfo{person}{Jun Sun}, \bibinfo{person}{Yun Lin}, {and}
  \bibinfo{person}{Quang~Tran Minh}.} \bibinfo{year}{2020}\natexlab{}.
\newblock \showarticletitle{SFuzz: An Efficient Adaptive Fuzzer for Solidity
  Smart Contracts}. In \bibinfo{booktitle}{\emph{Proceedings of the ACM/IEEE
  42nd International Conference on Software Engineering}}
  \emph{(\bibinfo{series}{ICSE '20})}. \bibinfo{publisher}{Association for
  Computing Machinery}, \bibinfo{address}{New York, NY, USA},
  \bibinfo{pages}{778–788}.
\newblock
\showISBNx{9781450371216}
\urldef\tempurl%
\url{https://doi.org/10.1145/3377811.3380334}
\showDOI{\tempurl}


\bibitem[\protect\citeauthoryear{Ni, Tian, Yang, Lo, Chen, and Yang}{Ni
  et~al\mbox{.}}{2023}]%
        {complexity}
\bibfield{author}{\bibinfo{person}{Chao Ni}, \bibinfo{person}{Cong Tian},
  \bibinfo{person}{Kaiwen Yang}, \bibinfo{person}{David Lo},
  \bibinfo{person}{Jiachi Chen}, {and} \bibinfo{person}{Xiaohu Yang}.}
  \bibinfo{year}{2023}\natexlab{}.
\newblock \showarticletitle{Automatic Identification of Crash-inducing Smart
  Contracts}. In \bibinfo{booktitle}{\emph{2023 IEEE International Conference
  on Software Analysis, Evolution and Reengineering (SANER)}}.
  \bibinfo{pages}{108--119}.
\newblock
\urldef\tempurl%
\url{https://doi.org/10.1109/SANER56733.2023.00020}
\showDOI{\tempurl}


\bibitem[\protect\citeauthoryear{Rodler, Paa{\ss}en, Li, Bernhard, Holz,
  Karame, and Davi}{Rodler et~al\mbox{.}}{2023}]%
        {rodler2023ef}
\bibfield{author}{\bibinfo{person}{Michael Rodler}, \bibinfo{person}{David
  Paa{\ss}en}, \bibinfo{person}{Wenting Li}, \bibinfo{person}{Lukas Bernhard},
  \bibinfo{person}{Thorsten Holz}, \bibinfo{person}{Ghassan Karame}, {and}
  \bibinfo{person}{Lucas Davi}.} \bibinfo{year}{2023}\natexlab{}.
\newblock \showarticletitle{EF/CF: High Performance Smart Contract Fuzzing for
  Exploit Generation}.
\newblock \bibinfo{journal}{\emph{arXiv preprint arXiv:2304.06341}}
  (\bibinfo{year}{2023}).
\newblock


\bibitem[\protect\citeauthoryear{smartbugs}{smartbugs}{2020}]%
        {8Smartbugs}
\bibfield{author}{\bibinfo{person}{smartbugs}.}
  \bibinfo{year}{2020}\natexlab{}.
\newblock \bibinfo{title}{Smartbugs wild dataset}.
\newblock
\newblock
\urldef\tempurl%
\url{https://github.com/smartbugs/smartbugs-wild}
\showURL{%
\tempurl}


\bibitem[\protect\citeauthoryear{So, Hong, and Oh}{So et~al\mbox{.}}{2021}]%
        {so2021smartest}
\bibfield{author}{\bibinfo{person}{Sunbeom So}, \bibinfo{person}{Seongjoon
  Hong}, {and} \bibinfo{person}{Hakjoo Oh}.} \bibinfo{year}{2021}\natexlab{}.
\newblock \showarticletitle{{SmarTest}: Effectively Hunting Vulnerable
  Transaction Sequences in Smart Contracts through Language Model-Guided
  Symbolic Execution}. In \bibinfo{booktitle}{\emph{30th USENIX Security
  Symposium (USENIX Security 21)}}. \bibinfo{pages}{1361--1378}.
\newblock


\bibitem[\protect\citeauthoryear{Su, Dai, Zhao, Zheng, and Luo}{Su
  et~al\mbox{.}}{2023}]%
        {su2022effectively}
\bibfield{author}{\bibinfo{person}{Jianzhong Su}, \bibinfo{person}{Hong-Ning
  Dai}, \bibinfo{person}{Lingjun Zhao}, \bibinfo{person}{Zibin Zheng}, {and}
  \bibinfo{person}{Xiapu Luo}.} \bibinfo{year}{2023}\natexlab{}.
\newblock \showarticletitle{Effectively Generating Vulnerable Transaction
  Sequences in Smart Contracts with Reinforcement Learning-Guided Fuzzing}. In
  \bibinfo{booktitle}{\emph{37th IEEE/ACM International Conference on Automated
  Software Engineering}} \emph{(\bibinfo{series}{ASE22})}.
  \bibinfo{publisher}{Association for Computing Machinery},
  \bibinfo{address}{New York, NY, USA}, Article \bibinfo{articleno}{36},
  \bibinfo{numpages}{12}~pages.
\newblock
\showISBNx{9781450394758}
\urldef\tempurl%
\url{https://doi.org/10.1145/3551349.3560429}
\showDOI{\tempurl}


\bibitem[\protect\citeauthoryear{Thummavet}{Thummavet}{2022a}]%
        {cross_function_Re}
\bibfield{author}{\bibinfo{person}{Phuwanai Thummavet}.}
  \bibinfo{year}{2022}\natexlab{a}.
\newblock \bibinfo{title}{Solidity Security By Example 04: Cross-Function
  Reentrancy}.
\newblock
\newblock
\urldef\tempurl%
\url{https://medium.com/valixconsulting/solidity-smart-contract-security-by-example-04-cross-function-reentrancy-de9cbce0558e}
\showURL{%
\tempurl}


\bibitem[\protect\citeauthoryear{Thummavet}{Thummavet}{2022b}]%
        {cross_contract_Re}
\bibfield{author}{\bibinfo{person}{Phuwanai Thummavet}.}
  \bibinfo{year}{2022}\natexlab{b}.
\newblock \bibinfo{title}{Solidity Security By Example 05: Cross-Contract
  Reentrancy}.
\newblock
\newblock
\urldef\tempurl%
\url{https://medium.com/valixconsulting/solidity-smart-contract-security-by-example-05-cross-contract-reentrancy-30f29e2a01b9}
\showURL{%
\tempurl}


\bibitem[\protect\citeauthoryear{Torres, Sch{\"u}tte, and State}{Torres
  et~al\mbox{.}}{2018}]%
        {13torres2018osiris}
\bibfield{author}{\bibinfo{person}{Christof~Ferreira Torres},
  \bibinfo{person}{Julian Sch{\"u}tte}, {and} \bibinfo{person}{Radu State}.}
  \bibinfo{year}{2018}\natexlab{}.
\newblock \showarticletitle{Osiris: Hunting for integer bugs in ethereum smart
  contracts}. In \bibinfo{booktitle}{\emph{Proceedings of the 34th Annual
  Computer Security Applications Conference}}. \bibinfo{pages}{664--676}.
\newblock


\bibitem[\protect\citeauthoryear{Tsankov, Dan, Drachsler-Cohen, Gervais,
  B\"{u}nzli, and Vechev}{Tsankov et~al\mbox{.}}{2018}]%
        {tsankov2018securify}
\bibfield{author}{\bibinfo{person}{Petar Tsankov}, \bibinfo{person}{Andrei
  Dan}, \bibinfo{person}{Dana Drachsler-Cohen}, \bibinfo{person}{Arthur
  Gervais}, \bibinfo{person}{Florian B\"{u}nzli}, {and} \bibinfo{person}{Martin
  Vechev}.} \bibinfo{year}{2018}\natexlab{}.
\newblock \showarticletitle{Securify: Practical Security Analysis of Smart
  Contracts}. In \bibinfo{booktitle}{\emph{Proceedings of the 2018 ACM SIGSAC
  Conference on Computer and Communications Security}}
  \emph{(\bibinfo{series}{CCS '18})}. \bibinfo{publisher}{Association for
  Computing Machinery}, \bibinfo{address}{New York, NY, USA},
  \bibinfo{pages}{67–82}.
\newblock
\showISBNx{9781450356930}
\urldef\tempurl%
\url{https://doi.org/10.1145/3243734.3243780}
\showDOI{\tempurl}


\bibitem[\protect\citeauthoryear{Witek~Radomski}{Witek~Radomski}{2018}]%
        {ERC1155}
\bibfield{author}{\bibinfo{person}{Philippe~Castonguay Witek~Radomski,
  Andrew~Cooke}.} \bibinfo{year}{2018}\natexlab{}.
\newblock \bibinfo{title}{ERC-1155: Multi Token Standard}.
\newblock
\newblock
\urldef\tempurl%
\url{https://eips.ethereum.org/EIPS/eip-1155}
\showURL{%
\tempurl}


\bibitem[\protect\citeauthoryear{Wood et~al\mbox{.}}{Wood
  et~al\mbox{.}}{2014}]%
        {wood2014ethereum}
\bibfield{author}{\bibinfo{person}{Gavin Wood} {et~al\mbox{.}}}
  \bibinfo{year}{2014}\natexlab{}.
\newblock \showarticletitle{Ethereum: A secure decentralised generalised
  transaction ledger}.
\newblock \bibinfo{journal}{\emph{Ethereum project yellow paper}}
  \bibinfo{volume}{151}, \bibinfo{number}{2014} (\bibinfo{year}{2014}),
  \bibinfo{pages}{1--32}.
\newblock


\bibitem[\protect\citeauthoryear{Ye, Ma, Lin, Sui, and Xue}{Ye
  et~al\mbox{.}}{2020}]%
        {11ye2020clairvoyance}
\bibfield{author}{\bibinfo{person}{Jiaming Ye}, \bibinfo{person}{Mingliang Ma},
  \bibinfo{person}{Yun Lin}, \bibinfo{person}{Yulei Sui}, {and}
  \bibinfo{person}{Yinxing Xue}.} \bibinfo{year}{2020}\natexlab{}.
\newblock \showarticletitle{Clairvoyance: Cross-contract static analysis for
  detecting practical reentrancy vulnerabilities in smart contracts}. In
  \bibinfo{booktitle}{\emph{Proceedings of the ACM/IEEE 42nd International
  Conference on Software Engineering: Companion Proceedings}}.
  \bibinfo{pages}{274--275}.
\newblock


\bibitem[\protect\citeauthoryear{Zhang, Zhang, Zhang, and Lin}{Zhang
  et~al\mbox{.}}{2020}]%
        {zhang2020txspector}
\bibfield{author}{\bibinfo{person}{Mengya Zhang}, \bibinfo{person}{Xiaokuan
  Zhang}, \bibinfo{person}{Yinqian Zhang}, {and} \bibinfo{person}{Zhiqiang
  Lin}.} \bibinfo{year}{2020}\natexlab{}.
\newblock \showarticletitle{TXSPECTOR: Uncovering Attacks in Ethereum from
  Transactions}. In \bibinfo{booktitle}{\emph{Proceedings of the 29th USENIX
  Conference on Security Symposium}} \emph{(\bibinfo{series}{SEC'20})}.
  \bibinfo{publisher}{USENIX Association}, \bibinfo{address}{USA}, Article
  \bibinfo{articleno}{156}, \bibinfo{numpages}{18}~pages.
\newblock
\showISBNx{978-1-939133-17-5}


\bibitem[\protect\citeauthoryear{Zhang, Banescu, Pasos, Stewart, and
  Ganesh}{Zhang et~al\mbox{.}}{2019}]%
        {24zhang2019mpro}
\bibfield{author}{\bibinfo{person}{William Zhang}, \bibinfo{person}{Sebastian
  Banescu}, \bibinfo{person}{Leonardo Pasos}, \bibinfo{person}{Steven Stewart},
  {and} \bibinfo{person}{Vijay Ganesh}.} \bibinfo{year}{2019}\natexlab{}.
\newblock \showarticletitle{Mpro: Combining static and symbolic analysis for
  scalable testing of smart contract}. In \bibinfo{booktitle}{\emph{2019 IEEE
  30th International Symposium on Software Reliability Engineering (ISSRE)}}.
  IEEE, \bibinfo{pages}{456--462}.
\newblock


\bibitem[\protect\citeauthoryear{Zheng, Zheng, and Luo}{Zheng
  et~al\mbox{.}}{2022}]%
        {zheng2022park}
\bibfield{author}{\bibinfo{person}{Peilin Zheng}, \bibinfo{person}{Zibin
  Zheng}, {and} \bibinfo{person}{Xiapu Luo}.} \bibinfo{year}{2022}\natexlab{}.
\newblock \showarticletitle{Park: Accelerating Smart Contract Vulnerability
  Detection via Parallel-Fork Symbolic Execution}. In
  \bibinfo{booktitle}{\emph{Proceedings of the 31st ACM SIGSOFT International
  Symposium on Software Testing and Analysis}} \emph{(\bibinfo{series}{ISSTA
  2022})}. \bibinfo{publisher}{Association for Computing Machinery},
  \bibinfo{address}{New York, NY, USA}, \bibinfo{pages}{740–751}.
\newblock
\showISBNx{9781450393799}
\urldef\tempurl%
\url{https://doi.org/10.1145/3533767.3534395}
\showDOI{\tempurl}


\bibitem[\protect\citeauthoryear{Zheng, Ning, Wang, Zhang, Zheng, Ye, and
  Chen}{Zheng et~al\mbox{.}}{2024}]%
        {zheng2024survey}
\bibfield{author}{\bibinfo{person}{Zibin Zheng}, \bibinfo{person}{Kaiwen Ning},
  \bibinfo{person}{Yanlin Wang}, \bibinfo{person}{Jingwen Zhang},
  \bibinfo{person}{Dewu Zheng}, \bibinfo{person}{Mingxi Ye}, {and}
  \bibinfo{person}{Jiachi Chen}.} \bibinfo{year}{2024}\natexlab{}.
\newblock \bibinfo{title}{A Survey of Large Language Models for Code:
  Evolution, Benchmarking, and Future Trends}.
\newblock
\newblock
\showeprint[arxiv]{cs.SE/2311.10372}


\bibitem[\protect\citeauthoryear{Zheng, Su, Chen, Lo, Zhong, and Ye}{Zheng
  et~al\mbox{.}}{2023a}]%
        {zheng2023dappscan}
\bibfield{author}{\bibinfo{person}{Zibin Zheng}, \bibinfo{person}{Jianzhong
  Su}, \bibinfo{person}{Jiachi Chen}, \bibinfo{person}{David Lo},
  \bibinfo{person}{Zhijie Zhong}, {and} \bibinfo{person}{Mingxi Ye}.}
  \bibinfo{year}{2023}\natexlab{a}.
\newblock \bibinfo{title}{DAppSCAN: Building Large-Scale Datasets for Smart
  Contract Weaknesses in DApp Projects}.
\newblock
\newblock
\showeprint[arxiv]{cs.SE/2305.08456}


\bibitem[\protect\citeauthoryear{Zheng, Zhang, Su, Zhong, Ye, and Chen}{Zheng
  et~al\mbox{.}}{2023b}]%
        {turntherudder}
\bibfield{author}{\bibinfo{person}{Zibin Zheng}, \bibinfo{person}{Neng Zhang},
  \bibinfo{person}{Jianzhong Su}, \bibinfo{person}{Zhijie Zhong},
  \bibinfo{person}{Mingxi Ye}, {and} \bibinfo{person}{Jiachi Chen}.}
  \bibinfo{year}{2023}\natexlab{b}.
\newblock \showarticletitle{Turn the Rudder: A Beacon of Reentrancy Detection
  for Smart Contracts on Ethereum}. In \bibinfo{booktitle}{\emph{Proceedings of
  the 45th International Conference on Software Engineering}}
  \emph{(\bibinfo{series}{ICSE '23})}. \bibinfo{publisher}{IEEE Press},
  \bibinfo{pages}{295–306}.
\newblock
\showISBNx{9781665457019}
\urldef\tempurl%
\url{https://doi.org/10.1109/ICSE48619.2023.00036}
\showDOI{\tempurl}


\bibitem[\protect\citeauthoryear{Zhou, Xiong, Ernstberger, Chaliasos, Wang,
  Wang, Qin, Wattenhofer, Song, and Gervais}{Zhou et~al\mbox{.}}{2023}]%
        {zhou2023sok}
\bibfield{author}{\bibinfo{person}{Liyi Zhou}, \bibinfo{person}{Xihan Xiong},
  \bibinfo{person}{Jens Ernstberger}, \bibinfo{person}{Stefanos Chaliasos},
  \bibinfo{person}{Zhipeng Wang}, \bibinfo{person}{Ye Wang},
  \bibinfo{person}{Kaihua Qin}, \bibinfo{person}{Roger Wattenhofer},
  \bibinfo{person}{Dawn Song}, {and} \bibinfo{person}{Arthur Gervais}.}
  \bibinfo{year}{2023}\natexlab{}.
\newblock \showarticletitle{Sok: Decentralized finance (defi) attacks}. In
  \bibinfo{booktitle}{\emph{2023 IEEE Symposium on Security and Privacy (SP)}}.
  IEEE, \bibinfo{pages}{2444--2461}.
\newblock


\bibitem[\protect\citeauthoryear{Zhou, Yang, Xiang, Cao, Yang, and Zhang}{Zhou
  et~al\mbox{.}}{2020}]%
        {zhou2020ever}
\bibfield{author}{\bibinfo{person}{Shunfan Zhou}, \bibinfo{person}{Zhemin
  Yang}, \bibinfo{person}{Jie Xiang}, \bibinfo{person}{Yinzhi Cao},
  \bibinfo{person}{Min Yang}, {and} \bibinfo{person}{Yuan Zhang}.}
  \bibinfo{year}{2020}\natexlab{}.
\newblock \showarticletitle{An Ever-Evolving Game: Evaluation of Real-World
  Attacks and Defenses in Ethereum Ecosystem}. In
  \bibinfo{booktitle}{\emph{Proceedings of the 29th USENIX Conference on
  Security Symposium}} \emph{(\bibinfo{series}{SEC'20})}.
  \bibinfo{publisher}{USENIX Association}, \bibinfo{address}{USA}, Article
  \bibinfo{articleno}{157}, \bibinfo{numpages}{17}~pages.
\newblock
\showISBNx{978-1-939133-17-5}


\end{thebibliography}
\nocite{*} %


\end{document}